\documentclass[sn-chicago]{sn-jnl}% Chicago-based Humanities Reference Style
\jyear{2023}%

\theoremstyle{thmstyleone}%
\theoremstyle{thmstyletwo}%

\theoremstyle{thmstylethree}%

\newcounter{remark_counter}

\usepackage{siunitx}

\newcommand{\smolhalf}{\textstyle\frac{1}{2}}

\begin{document}

\title[Freeform design with algorithmic differentiable ray tracing]{Gradient descent-based freeform optics design using algorithmic differentiable non-sequential ray tracing}

%%=============================================================%%
%% Prefix	-> \pfx{Dr}
%% GivenName	-> \fnm{Joergen W.}
%% Particle	-> \spfx{van der} -> surname prefix
%% FamilyName	-> \sur{Ploeg}
%% Suffix	-> \sfx{IV}
%% NatureName	-> \tanm{Poet Laureate} -> Title after name
%% Degrees	-> \dgr{MSc, PhD}
%% \author*[1,2]{\pfx{Dr} \fnm{Joergen W.} \spfx{van der} \sur{Ploeg} \sfx{IV} \tanm{Poet Laureate} 
%%                 \dgr{MSc, PhD}}\email{iauthor@gmail.com}
%%=============================================================%%

\author*[1,2]{\fnm{Bart} \sur{de Koning} \dgr{}}\email{B.deKoning@tudelft.nl}

\author[2]{\fnm{Alexander} \sur{Heemels} \dgr{}}\email{A.N.M.Heemels@tudelft.nl}
%\equalcont{These authors contributed equally to this work.}

\author[2]{\pfx{} \fnm{Aur\`ele} \sur{Adam}}\email{A.J.L.Adam@tudelft.nl}

\author[1]{\pfx{} \fnm{Matthias} \sur{M\"oller}}\email{M.Moller@tudelft.nl}

\affil[1]{\orgdiv{Numerical Analysis}, \orgname{Delft University of Technology}, \orgaddress{\street{Mekelweg 4}, \city{Delft}, \postcode{2628 CD}, \state{Zuid-Holland}, \country{Netherlands}}}

\affil[2]{\orgdiv{Applied Physics}, \orgname{Delft University of Technology}, \orgaddress{\street{Lorentzweg 1}, \city{Delft}, \postcode{2628 CJ}, \state{Zuid-Holland}, \country{Netherlands}}}

\makeatletter
\renewcommand*\env@matrix[1][\arraystretch]{%
  \edef\arraystretch{#1}%
  \hskip -\arraycolsep
  \let\@ifnextchar\new@ifnextchar
  \array{*\c@MaxMatrixCols c}}
\makeatother

%%==================================%%
%% sample for unstructured abstract %%
%%==================================%%

\abstract{
\emph{Algorithmic differentiable ray tracing} is a new paradigm that allows one to
solve the forward problem of how light propagates through an optical system while obtaining gradients of the simulation results with respect to parameters specifying the optical system. 
Specifically, the use of algorithmically differentiable non-sequential ray tracing provides an opportunity in the field of illumination design.
We demonstrate its potential by designing freeform lenses that project a prescribed irradiance distribution onto a plane. 
The challenge consists in finding a suitable surface geometry of the lens so that the light emitted by a light source is redistributed into a desired irradiance distribution. 
We discuss the crucial steps allowing the non-sequential ray tracer to be differentiable.
The obtained gradients are used to optimize the geometry of the freeform, and we investigate the effectiveness of adding a multi-layer perceptron neural network to the optimization that outputs parameters defining the freeform lens.
Lenses are designed for various sources such as collimated ray bundles or point sources, and finally, a grid of point sources approximating an extended source. 
The obtained lens designs are finally validated using the commercial non-sequential ray tracer LightTools.
}


%%\pacs[JEL Classification]{D8, H51}

%%\pacs[MSC Classification]{35A01, 65L10, 65L12, 65L20, 65L70}

\maketitle
\newpage
\section{Introduction}\label{sec:Introduction}
In the field of illumination optics, optical engineers design optical elements to transport the light from a source, which can be an LED, laser, or incandescent lamp, to obtain a desired irradiance (spatial density of the luminous flux) or intensity (angular density of the luminous flux) \citep{Grant2011}.
To transport the light from the source to the target, the optical engineer can construct a system consisting of various optical elements such as lenses, mirrors, diffusers, and light guides \citep{John2013}.
One particular type of optic used in automotive and road lighting applications is the freeform lens, a lens without any form of symmetry \citep{Falaggis2022, Mohedano2016}.
The design of these lenses is a complex problem. It is currently solved by numerically solving system-specific differential equations or through optimization, with every step validated using a (non-differentiable) non-sequential ray tracer \citep{Wu2018}.
Great effort is involved in generalizing these methods to account for varying amounts of optical surfaces \citep{Anthonissen2021}, their optical surface and volume properties \citep{Kronberg2022, lippman_prescribed_2020}, or the source model~\citep{Muschaweck2022, tukker_efficient_2007, sorgato_design_2019}. 

The performance of an optical system is evaluated using ray tracing, which is the process of calculating the path of a ray originating from a source through the optical system. Sequential ray tracers such as Zemax~\citep{zemax} and Code V~\citep{codev}, primarily used in the design of imaging optics, trace a small number of rays to determine the quality of the image. 
Non-sequential ray tracers such as LightTools~\citep{LightTools} and Photopia~\citep{photopia} use many rays to simulate the optical flux through the system and share similarities with the rendering procedures in computer graphics, with the main difference being that the rays are traced from source to camera.

Algorithmically differentiable ray tracing, a generalization of differential ray tracing \citep{feder_differentiation_1968, stone_differential_1997, oertmann_differential_1989, chen_second-order_2012}, is a tool that is being developed for both sequential~\citep{Sun2021, Volatier2017}  and non-sequential \citep{Mitsuba2} ray tracing.
\emph{Differential ray tracing} obtains system parameter gradients using numerical or algebraic differentiation. The gradient can be calculated numerically using numerical differentiation or the adjoint method \citep{givoli_tutorial_2021}, requiring the system to be ray traced twice, once for its current state and once with perturbed system parameters. Analytic expressions for the gradient can be obtained by tracing the rays analytically through the system, calculating where the ray intersects the system's surfaces and how the ray's trajectory is altered. However, these expressions can become long and complicated depending on the system. In addition, the method is limited to optics described by conics as finding analytic ray surface intersection with surfaces of higher degrees becomes complicated or even impossible. Algorithmic differentiable ray tracing can handle these issues by obtaining the gradients with one single forward simulation for an almost arbitrary system. In addition, it can be seamlessly integrated into gradient-descent-based optimization pipelines. A modern framework for this is \emph{Physics Informed Machine Learning} \citep{Karniadakis2021}, where a neural network is trained to approximate the solution to a physics problem formulated using data, a set of differential equations, or an implemented physics simulation (or a combination of these). 

We investigate the reliability of designing freeform lenses with B-spline surfaces \citep{Piegl1995} using algorithmically differentiable non-sequential ray tracing and gradient-based optimization to redirect the light of a light source into a prescribed irradiance distribution. The source models will be the collimated light source, point source, and finally, sources with a finite extent. The results are validated using the commercial ray trace program LightTools \citep{LightTools}. In addition, we investigate the effectiveness of optimizing a network to determine the optimal B-spline control points as proposed in \citep{Moller2021PIML} and \citep{GASICK2023115839}, and compare it to optimizing the control points directly and seeing the possible speed-up.

\section{Gradient-based freeform design}
\begin{figure}
    \centering
    \includegraphics[width = \textwidth]{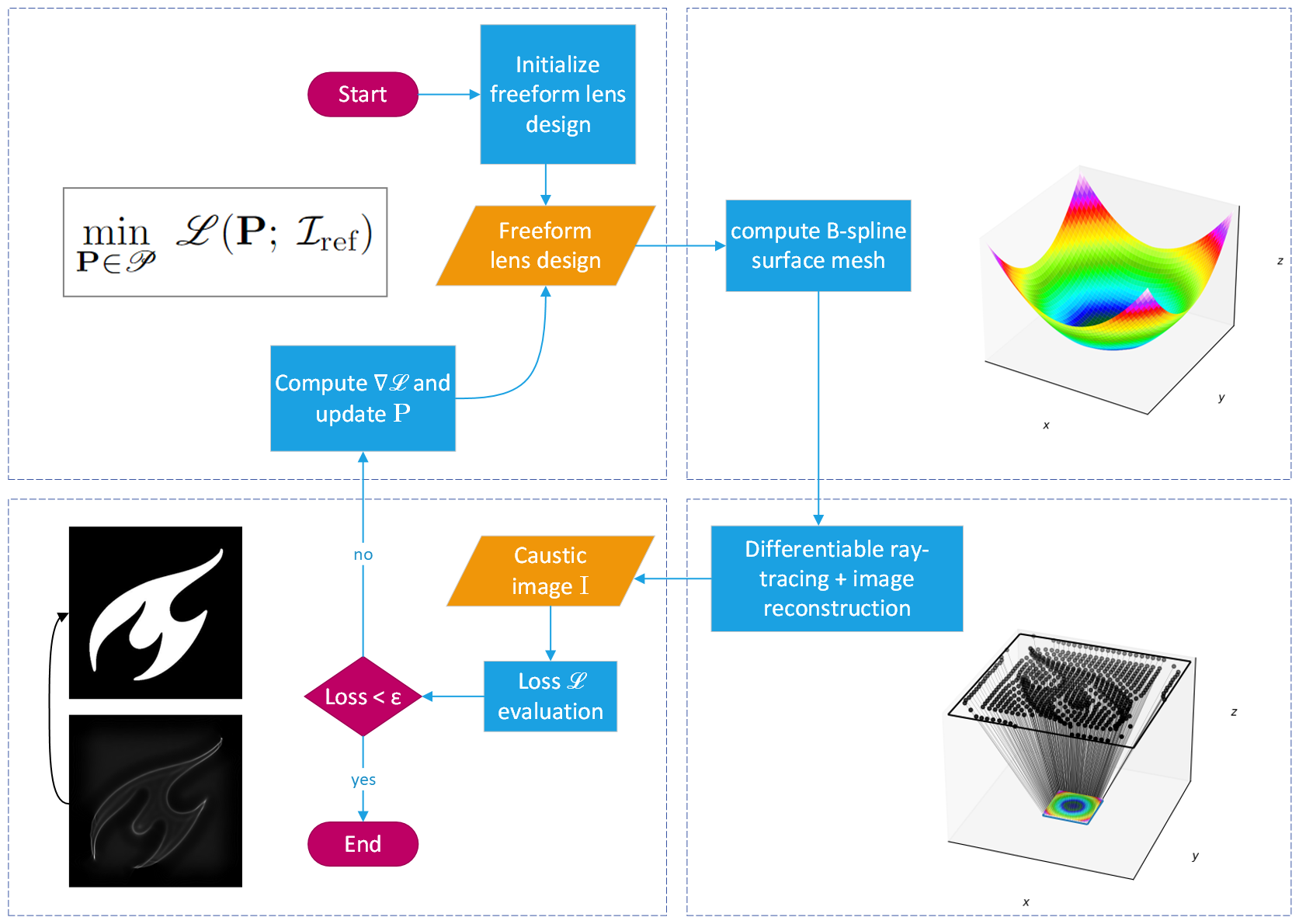}
    \caption{Overview of our learning-based freeform design pipeline.}
    \label{fig:optimizationloop}
\end{figure}

The overall structure of our pipeline is depicted in Fig.~\ref{fig:optimizationloop}. 
A freeform surface is defined by the parameters $P \in \mathscr{P}$, where $\mathscr{P}$ is the set of permissible parameter values. This surface is combined with a flat surface to create a lens, and an irradiance distribution $\mathcal{I}$ is produced by tracing rays through the lens onto a screen. The irradiance distribution is compared to a target $\mathcal{I}_\text{ref}$ yielding a loss $\mathscr{L}(\mathbf{P};\mathcal{I}_\text{ref})$. The optimization problem we are trying to solve can then be formulated as
\begin{equation}
    \min_\mathbf{P \in \mathscr{P}} \; \mathscr{L}(\mathbf{P};\mathcal{I}_\text{ref}),
\end{equation}
which we solve by using gradient descent.

The freeform surface of the lens is defined in terms of a B-spline surface. From a manufacturing standpoint, this is convenient since B-spline surfaces can be chosen to be $C^1$ smooth (in fact, B-spline surfaces can be $C^n$ smooth for arbitrarily large $n$). From an optimization perspective, B-spline surfaces have the property that the control points that govern the shape of the surface and which will be optimized have a local influence on the surface geometry, which in turn has a local influence on the resulting irradiance distribution.

\subsection{The lens model using a B-spline surface}
\begin{figure}
    \centering
    \includegraphics[width = 0.2\textwidth]{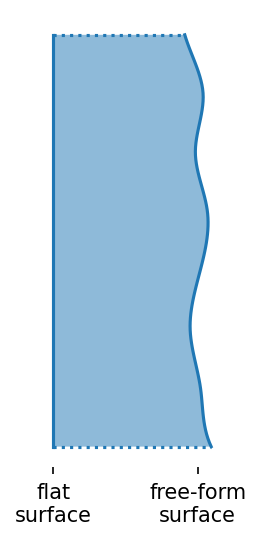}
    \caption{The used lens type: a volume enclosed between a flat surface and a freeform surface with a uniform refractive index.}
    \label{fig:lenschematic}
\end{figure}

We define a lens as in Fig.~\ref{fig:lenschematic} as the volume between a flat surface and a B-spline surface, with a uniform refractive index.

\begin{figure}
    \centering
    \includegraphics[width = 0.5\textwidth]{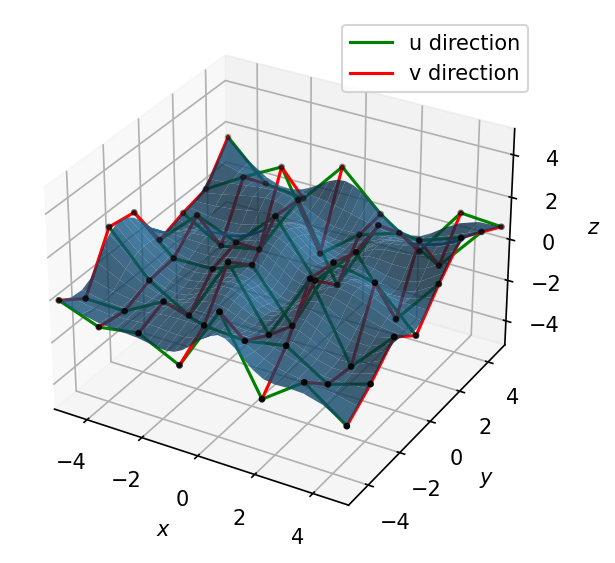}
    \caption{A B-spline surface of degrees $(p,q) = (3,3)$ with indicated directions of the $u$ and $v$ parameters. The control points are shown in black.}
    \label{fig:Bsplinesurface}
\end{figure}

A B-spline surface $\mathbf{S}$ in $\mathbb{R}^3$ is a parametric surface, see Fig.~\ref{fig:Bsplinesurface}. It has rectangular support $[a,b]\times[c,d]$ where $a<b$ and $c<d$. It is defined as a linear combination of an $(n_1+1) \times (n_2+1)$ grid of control points $\mathbf{P}_{i,j} \in \mathbb{R}^3$ where $n_1$ and $n_2$ are positive integers which define the size of the control net $\{\mathbf{P}_{i,j} :  0 \leq i \leq n_1, \; 0 \leq j \leq n_2\}$. This linear combination is defined in terms of univariate B-spline basis functions $N_{i,p}$ and $N_{j,q}$, which are recursively defined as follows by the Cox-de Boor formula \cite[eq. 2.5]{Piegl1995}:
\begin{align}
    & N_{i,0}(u) =
    \begin{cases}
        1 \text{ if } u_i \le u < u_{i+1} \\
        0 \text{ otherwise}
    \end{cases} \nonumber \\
    & N_{i,p}(u) = \frac{u-u_i}{u_{i+p}-u_i}N_{i,p-1}(u) + \frac{u_{i+p+1}-u}{u_{i+p+1}}N_{i+1,p-1}(u),
\end{align}
for $i=0,\ldots,n_1$ and similarly for $N_{j,q}$ with $j=0,\ldots,n_2$. The basis functions are $p$-degree piece-wise polynomials. The knots $u_i$ are non-decreasing real numbers in $[a,b]$ collected in a knot vector
\cite[eq. 2.13]{Piegl1995}:
\begin{equation}
    \mathcal{V} = (\underbrace{0,\ldots,0}_{p+1},u_{p+1},\ldots,u_{r-p-1},\underbrace{1,\ldots,1}_{p+1}) \in \mathbb{R}^{r+1}.
\end{equation}
 Here we assume $a=0$ and $b=1$. For the number of knots $r+1$, the number of control points $n_1+1$ and the degree $p$, there is the relationship $n_1 = r-p-1$. 

An interval $[u_i, u_{i+1})$ of consecutive knots is called a knot span, on which the basis functions are analytic. At the knots, the basis functions are $p-k$ times continuously differentiable, with $k$ being the multiplicity of the knot, that is, how many times the same knot is repeated in the knot vector. The specific type of knot vector here with the multiplicity of $p+1$ at the first and last knots is called clamped or open, which yields the property
\begin{equation}
    N_{0,p}(0) = N_{n_1,p}(1) = 1.
\end{equation}
All other basis functions are $0$ at $u=0,1$, which means that the boundary control points determine the boundary of the surface.  

\noindent The interior knots are chosen to be equispaced, \emph{i.e.}
\begin{equation}
    u_i = \frac{i-p}{n_1-p+1}, \qquad i = p,\ldots, n_1+1.
\end{equation}
Thus there are no knots with multiplicity larger than $1$ apart from those at the boundary. So if $p,q \ge 2$, then the surface is at least $C^1$, and the gradient of the B-spline surface is defined everywhere.

Completely analogously, there is the knot vector $\mathcal{W}$ with the knots $v_j$ for the basis functions $N_{j,q}$  of degree $q$, for $j=0,\ldots,n_2$. The definition of the surface is then \cite[eq. 3.11]{Piegl1995}
\begin{equation}
    \mathbf{S}(u,v) = \sum_{i=0}^{n_1}\sum_{j=0}^{n_2} N_{i,p}(u)N_{j,q}(v)\mathbf{P}_{i,j}, \quad (u,v) \in [0,1]^2. \label{eq:Bsplinesurfacedef}
\end{equation}

\subsubsection{Linearizing the B-spline parametrizations}
The volume $V \subset \mathbb{R}^3$ of the modeled lens has a rectangular extent  $[-r_x,r_x] \times [-r_y,r_y]$ in the ($x,y$)-plane with $2r_x$ and $2r_y$ being the width and height of the lens, respectively. The lens volume is enclosed on one side by a B-spline surface
\begin{equation}
    \mathbf{S}(u,v) = (X(u,v), Y(u,v), Z(u,v)), \label{eq:coordfuncs}
\end{equation}
where $X,Y,Z$ are the individual coordinate parameterizations, for instance
\begin{subequations}
    \begin{align}
        X(u,v) = \sum_{i=0}^{n_1}\sum_{j=0}^{n_2} N_{i,p}(u)N_{j,q}(v)P^x_{i,j}, \\
        Y(u,v) = \sum_{i=0}^{n_1}\sum_{j=0}^{n_2} N_{i,p}(u)N_{j,q}(v)P^x_{i,j}.
    \end{align}
\end{subequations}
For simplicity of calculations on ray-sampling and ray-intersection (Section \ref{subsec:raytracer}), it is helpful to define the mapping $(u,v) \mapsto (X(u,v),Y(u,v))$ in a way that it is analytically invertible. Therefore the coordinates of the control points are chosen such that the parametrizations $X$ and $Y$ are linear:
\begin{subequations}
    \begin{align}
        X &: u \mapsto (2u-1)r_x \in [-r_x,r_x], \label{eq:linearX}\\
        Y &: v \mapsto (2v-1)r_y \in [-r_y,r_y]. \label{eq:linearY}
    \end{align}
\end{subequations}
In general, $X$ and $Y$ are degree $p$ and $q$ piece-wise polynomials, respectively, and thus not linear. Linearity can be achieved by making use of the nodal representation of the B-spline 
 basis functions \cite[eq. 23]{Cohen2010Iso}:
\begin{equation}
    u = \sum_{i=0}^{n_1} u_{i,p}^* N_{i,p}(u), \quad u \in [0,1], \quad u_{i,p}^* = \frac{u_{i+1}+\ldots+u_{i+p}}{p}, \label{eq:nodalrepresentation}
\end{equation}
which provides a specific knot vector-dependent linear combination of the basis functions that yields the identity function on the domain $[0,1]$. The values $u_{i,p}^*$ are called the Greville abscissae \cite[sec. 8.6]{Farin2002}.

We assume that the $P^x_{i,j}$ are independent of $j$, and choose $j=0$ as a representative.  Then we obtain by the definition of $X$:
\begin{subequations}
    \begin{align}
        X(u,v) &= \sum_{i=0}^{n_1}\sum_{j=0}^{n_2} P^x_{i,j}N_{i,p}(u)N_{j,q}(v) \\
        &= \sum_{i=0}^{n_1} P^x_{i,0}N_{i,p}(u) \underbrace{\sum_{j=0}^{n_2} N_{j,q}(v)}_{=1},
    \end{align}
\end{subequations}
where $\sum_{i=0}^n N_{i,p}(u) = 1$ by the partition of unity property of the basis functions \cite[P2.4]{Piegl1995}. Now we see that if we define $P^x_{i,j} := u^*_{i,p}$ then $X(u) = u$. Thus if we apply the mapping $u \mapsto (2u-1)r_x$ to both sides of eq. ~\ref{eq:nodalrepresentation}, we obtain
\begin{equation}
    (2u-1)r_x = 
    \sum_{i=0}^{n_1} (2u^*_{i,p}-1)r_x N_{i,p}(u). \label{eq:lineartaucontrolpoints}
\end{equation}
This equality can be understood by expanding the $1$ into the sum over all $N_{i,p}(u)$ by again exploiting the partition of unity property. Thus if we define $P^x_{i,j} := (2u^*_{i,p}-1)r_x$ and equivalently $P^y_{i,j} := (2v^*_{j,q}-1)r_y$, then Eqs.~\ref{eq:linearX} and \ref{eq:linearY} and are satisfied.

The lens is then defined as the volume in $\mathbb{R}^3$ enclosed by the B-spline surface $\mathbf{S}$ and the flat surface given by $z=z_\text{in}$ on $[-r_x,r_x] \times [-r_y,r_y]$:
\begin{equation}
    V = \left\{ (x,y,z) \in \mathbb{R}^3 \; \mid \;
    z_\text{in} \le z \le 
    Z\left(X^{-1}(x), Y^{-1}(y)\right), \vert x\vert\le r_x, \vert y\vert\le r_y\right\}.
\end{equation}
For the arguments of $Z(u,v)$ the inverses of $X$ and $Y$ are used:
\begin{equation}
    X^{-1}(x) = \frac{1}{2}\left(\frac{x}{r_x}+1\right), \quad
    Y^{-1}(y) = \frac{1}{2}\left(\frac{y}{r_y}+1\right).
\end{equation}

\subsubsection{Lens constraints} \label{subsubsec:lensconstraints}
To let the lens be well-defined the surfaces of the lens should not intersect:
\begin{equation}
    z_\text{in} < Z(u,v), \quad(u,v) \in [0,1]^2.
\end{equation}
By the convex hull property of B-spline surfaces \cite[P3.22]{Piegl1995} it suffices that
\begin{equation}
    P^z_{i,j} > z_\text{in} \quad \forall (i,j). \label{eq:nosurfintersect}
\end{equation}
Manufacturing can require that the lens has some minimal thickness $\delta$, so that the constraint is stronger:
\begin{equation}
    P^z_{i,j} \ge \delta + z_\text{in} \quad \forall (i,j).
\end{equation}

\subsection{Differentiable ray tracer} \label{subsec:raytracer}
Our implementation traces rays from a source through the flat lens surface and the freeform lens surface to the detector screen as depicted in Figs.~\ref{fig:planetraceschematic} and \ref{fig:pointtraceschematic}. Other ray paths, e.g., total internal reflection at lens surfaces, are not considered since it is assumed that the contribution of these to the resulting irradiance distribution is negligible.

\subsubsection{Sources and ray-sampling}
\begin{figure}
    \centering
    \includegraphics[width = 0.6\textwidth]{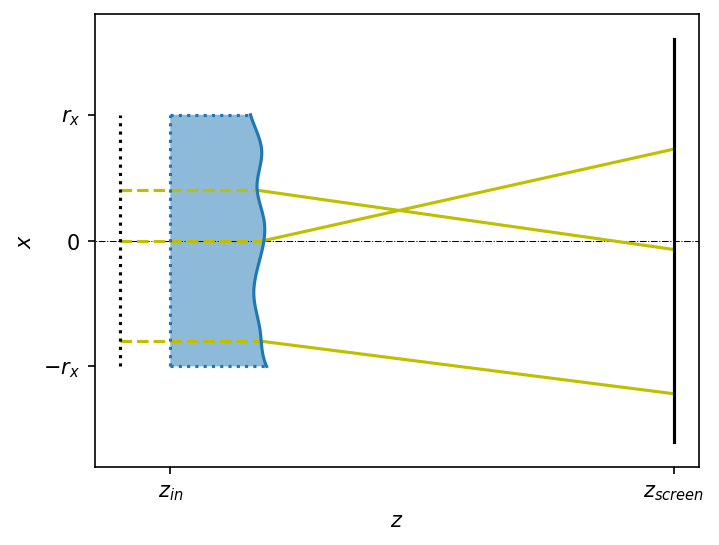}
    \caption{Schematic of the ray tracing with a collimated ray bundle source.}
    \label{fig:planetraceschematic}
\end{figure}

\begin{figure}
    \centering
    \includegraphics[width = 0.6\textwidth]{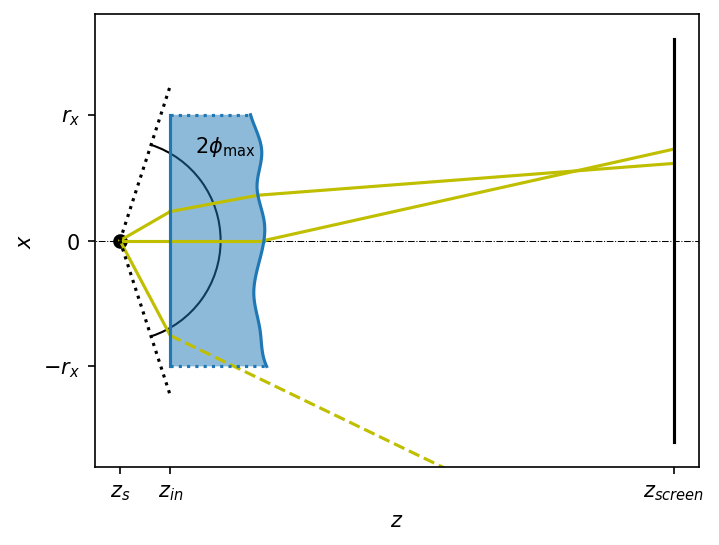}
    \caption{Schematic of the ray tracing with a point source.}
    \label{fig:pointtraceschematic}
\end{figure}

Non-sequential ray tracing is a Monte-Carlo approximation method of the solution to the continuous integration formulation of light transport through an optical system. For a detailed discussion of this topic, see \cite[ch. 14]{pharr2016physically}. Thus to perform ray tracing, the light emitted by a source must be discretized into a finite set of rays
\begin{equation}
    l: t \rightarrow \mathbf{o} + \hat{\mathbf{d}}t,
\end{equation}
where $\mathbf{o}$ is the origin of the ray and $\hat{\mathbf{d}}$ its normalized direction vector. Both collimated ray bundle and point sources will be considered, see Figs.~\ref{fig:planetraceschematic} and \ref{fig:pointtraceschematic}, respectively. 

Tracing rays from a collimated ray bundle can be understood from Fig.~\ref{fig:planetraceschematic}. The path of all rays from the source plane to the B-spline surface is a line segment parallel to the $z$-axis. Therefore, we can sample the incoming rays directly on the B-spline surface, with $\hat{\mathbf{d}} = (0,0,1)^\top$. By the linearity of $X$ and $Y$ sampling on the B-spline domain $[0,1]^2$ is analogous to sampling on the lens extent $[-r_x,r_x]\times [-r_y,r_y]$ in terms of distribution. Rays are sampled in a (deterministic) square grid on $[0,1]^2$. 

For a point source, each ray starts at the location of the source, and the direction vector $\hat{\mathbf{d}}$ is sampled over the unit sphere $\mathbb{S}^2$. More precisely, $\hat{\mathbf{d}}$ is given by
\begin{equation}
    \hat{\mathbf{d}} = \left(\cos\theta\sin\phi,\sin\theta\sin\phi,\cos\phi\right)^\top,
\end{equation}
with $\theta \in [0,2\pi)$ and $\phi \in [0,\phi_\text{max}]$ for some $0\le \phi_\text{max} < \frac{\pi}{2}$, see Fig.~\ref{fig:pointtraceschematic}. $\phi_\text{max}$ is chosen as small enough to minimize the number of rays that miss the lens entrance surface but large enough such that the whole surface is illuminated. For instance, if the source is on the $z$-axis, then $\phi_\text{max} = \arctan\left(\frac{\sqrt{r_x^2 + r_y^2}}{z_\text{in}-z_s}\right)$ where $z_\text{in}$ is the $z$-coordinate location of the entrance surface and $z_s$ the $z$-coordinate of the source. To uniformly sample points on this sphere segment, $\theta$ is sampled (non-deterministically) uniformly in $[0,2\pi)$ and $\phi$ is given by
\begin{equation}
    \phi = \arccos\left(1-(1-\cos\phi_\text{max})a\right)
\end{equation}
where $a$ is sampled (non-deterministically) uniformly in $[0,1]$. This sampling is used to produce the results in Section~\ref{sec:results}.

For the point source, the calculation of the intersection of a ray with the B-spline surface is non-trivial. This calculation comes down to finding the smallest positive root of the $p+q$ degree piece-wise polynomial function
\begin{equation}
    f(t) = 
    Z\left(\begin{pmatrix}o_u \\ o_v\end{pmatrix} + 
    \begin{pmatrix}d_u \\ d_v\end{pmatrix}t
    \right)
     - d_zt - o_z, \label{eq:surfaceintersect}
\end{equation}
if such a root exists and yields a point in the domain of $Z$.  Here the subscripts $u$ and $v$ denote that the ray is considered in $(u,v,z)$ space instead of $(x,y,z)$ space, so for instance
\begin{equation}
    o_u = X^{-1}(o_x) = \frac{1}{2}\left(\frac{o_x}{r_y}+1\right), \quad d_v = \frac{d_y}{2r_y}.
\end{equation}
The roots of eq. \ref{eq:surfaceintersect} cannot generally be found analytically for $p+q>4$, and thus an intersection algorithm is implemented, which is explained in the next section.

\subsubsection{B-spline surface intersection algorithm}
The intersection algorithm is based on constructing a triangle mesh approximation of the B-spline surface and computing intersections with that mesh.

\paragraph{Triangle mesh intersection phase 1: bounding boxes}
\begin{figure}
    \centering
    \includegraphics[width=0.6\textwidth]{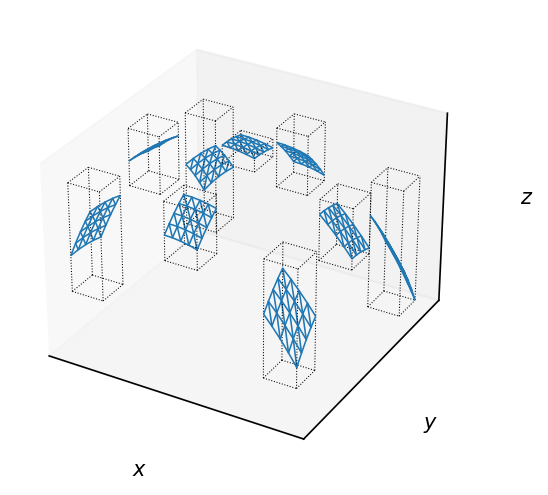}
    \caption{Triangles and corresponding bounding box for a few knot span products of a spherical surface.}
    \label{fig:bb_per_knotspanproduct}
\end{figure}
\begin{figure}
    \centering
    \includegraphics[width=0.6\textwidth]{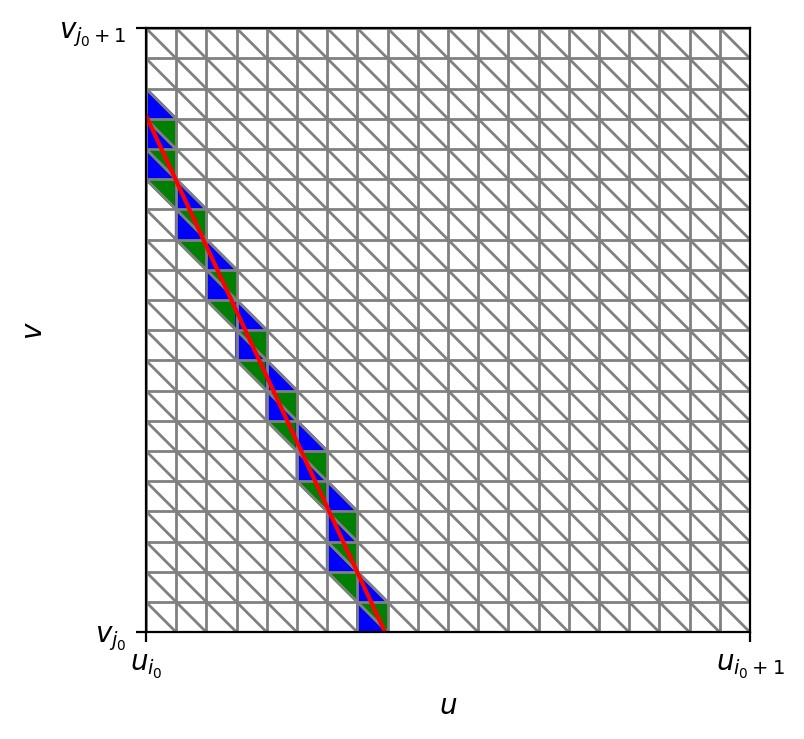}
    \caption{Example of which triangles are candidates for a ray-surface intersection with the ray plotted in red, based on their $u,v$-domain.}
    \label{fig:uvtriangleint}
\end{figure}
Checking every ray against every triangle for intersection is computationally expensive, so it is helpful to have bounding box tests that provide rough information about whether the ray is even near some section of the B-spline surface. B-spline theory provides a tool for this: the strong convex hull property, which yields the bounding box 

\begin{equation}
    B_{i_0,j_0} = \left[u_{i_0},u_{i_0 + 1}\right) \times \left[v_{j_0},v_{j_0 + 1}\right) \times \left[z^{\min}_{i_0,j_0}, z^{\max}_{i_0,j_0}\right]
\end{equation}
where $z^{\min}_{i,j}$ and $z^{\max}_{i,j}$ are the minimum and maximum $z$-values of the control points that affect the B-spline surface on the knot span product $\left[u_{i_0},u_{i_0 + 1}\right) \times \left[v_{j_0},u_{j_0 + 1}\right)$, hence those with indices $i_0-p\le i\le i_0, j_0-q \le j\le j_0$. Formulated in terms of $Z(u,v)$ this yields
\begin{equation}
    z^{\min}_{i_0,j_0} \le Z(u,v) \le z^{\max}_{i_0,j_0}, \quad
    (u,v) \in \left[u_{i_0},u_{i_0 + 1}\right) \times \left[v_{j_0},v_{j_0 + 1}\right).
\end{equation}
Examples of such bounding boxes are shown in Fig.~\ref{fig:bb_per_knotspanproduct}.

There are two steps in applying the bounding boxes in the intersection algorithm. First, a test for the entire surface (in $(u,v,z)$-space):
\begin{equation}
    [0,1]^2 \times \left[\min_{i,j}P^z_{i,j},\max_{i,j}P^z_{i,j}\right].
\end{equation}
Second, a recursive method where, starting with all knot span products, each rectangle of knot span products is divided into at most 4 sub-rectangles for a new bounding box test until individual knot span products are reached.

% Note that these bounding boxes make use of the strong convex hull property in a fairly weak way. A future iteration of this algorithm could make full use of the strong convex hull property to make the algorithm more efficient.

\paragraph{Triangle mesh intersection phase 2: $(u,v)$-space triangle intersection}
Each non-trivial knot span product $[u_{i_0},u_{i_0+1}) \times [v_{j_0},v_{j_0+1})$ is divided into a grid of $n_u$ by $n_v$ rectangles. Thus we can define the boundary points
\begin{subequations}
    \begin{align}
         u_{i_0,k} =& u_{i_0} + k\Delta u_{i_0},
         \quad \Delta u_{i_0} = \frac{u_{i_0+1}-u_{i_0}}{n_u}, 
         \quad k = 0, \ldots, n_u, \\
         v_{i_0,\ell} =&  v_{j_0}+ \ell \Delta v_{j_0}, \quad \Delta v_{j_0} = \frac{v_{j_0+1}-v_{j_0}}{n_v},
         \quad \ell = 0,\ldots, n_v.
     \end{align}
\end{subequations}

Each rectangle is divided into a lower left and an upper right triangle, as demonstrated in Fig.~\ref{fig:uvtriangleint}. In this figure it is shown for a ray projected onto the $(u,v)$-plane in some knot span which triangles are candidates for an intersection in $(u,v,z)$-space. This is determined by the following rules:
\begin{itemize}
    \item A lower left triangle is intersected in the $(u,v)$-plane if either its left or lower boundary is intersected by the ray;
    \item an upper right triangle is intersected in the $(u,v)$-plane if either its right or upper boundary is intersected by the ray.
\end{itemize}

The intersection of these boundaries can be determined by finding the indices of the horizontal lines at which the vertical lines are intersected:
\begin{equation}
    \ell_k = \left\lfloor\frac{ o_v+(u_{i_0,k}-o_u)\frac{d_v}{d_u} - v_{j_0}}{\Delta v_{j_0}}\right\rfloor,
\end{equation}
and analogously $k_\ell$.

\paragraph{Triangle mesh intersection phase 3: $u,v,z$-space triangle intersection}
A lower left triangle can be expressed by a plane
\begin{equation}
    T(u,v) = Au + Bv + C
\end{equation}
    defined by the following linear system:
\begin{equation}
    \begin{pmatrix}
    u_{i_0,k} & v_{j_0,\ell} & 1 \\
    u_{i_0,k+1} & v_{j_0,\ell} & 1 \\
    u_{i_0,k} & v_{j_0,\ell+1} & 1
    \end{pmatrix}
    \begin{pmatrix}
    A \\ B \\ C
    \end{pmatrix}
    =
    \begin{pmatrix}[1.75]
    z_{i_0,k}^{j_0,\ell} \\ z_{i_0,k+1}^{j_0,\ell} \\ z_{i_0,k}^{j_0,\ell+1}
    \end{pmatrix}.
\end{equation}
Here we use the following definition:
\begin{equation}
    z_{i_0,k}^{j_0,\ell} = Z(u_{i_0,k},v_{j_0,\ell}).
\end{equation}
This yields the plane
\begin{align}
    T(u,v) =&& z_{i_0,k}^{j_0,\ell} + n_u\left(z_{i_0,k+1}^{j_0,\ell}-z_{i_0,k}^{j_0,\ell}\right)\frac{u-u_{i_0,k}}{u_{i_0+1}-u_{i_0}} \\ 
    && +n_v\left(z_{i_0,k}^{j_0,\ell+1}-z_{i_0,k}^{j_0,\ell}\right)\frac{v-v_{j_0,\ell}}{v_{j_0+1}-v_{j_0}}. \label{eq:trianglefuncdetermined}
\end{align}
Note that to define this triangle, the B-spline basis functions are evaluated at fixed points in $[0,1]^2$ independent of the rays or the $P^z_{i,j}$. This means that for a lens that will be optimized these basis function values can be evaluated and stored only once rather than in every iteration, for computational efficiency.

Computing the intersection with the ray $\tilde{\mathbf{r}}(t) = \tilde{\mathbf{o}} + \tilde{\hat{\mathbf{d}}}t$ is now straight-forward, and yields
\begin{equation}
    t_\text{int} = - \frac{C+\langle\tilde{\mathbf{o}}, \mathbf{n}\rangle}{\langle\tilde{\hat{\mathbf{d}}},\mathbf{n}\rangle}, \quad \mathbf{n} = 
    \begin{pmatrix} 0 \\ 1 \\ \partial_u T\end{pmatrix} \times
    \begin{pmatrix} 1 \\ 0 \\ \partial_v T\end{pmatrix} = 
    \begin{pmatrix}A \\ B \\ -1\end{pmatrix},
\end{equation}
where $\mathbf{n}$ is a normal vector to the triangle, computed using the cross product. This also explains why $\langle\tilde{\hat{\mathbf{d}}},\mathbf{n}\rangle=0$ does not yield a well-defined result: in this situation the ray is parallel to the triangle.

The last thing to check is whether $\tilde{l}(t_\text{int})$ lies in the $(u,v)$-domain of the triangle, which can be checked by three inequalities for the three boundaries of the triangle:

\begin{subequations}
    \begin{align}
        o_u + d_u t_\text{int} \ge u_{i_0,k} \\
        0 \leq o_v + d_v t_\text{int} - v_{j_0,\ell}< \frac{n_u}{n_v}\frac{v_{j_0+1}-v_{j_0}}{u_{i_0+1}-u_{i_0}}(u_{i_0,k+1}-(o_u + d_u t_\text{int})).
    \end{align}
\end{subequations}

The computation for an upper right triangle is completely analogous. The upper triangle has a closed boundary, whereas the lower triangle has an open one and vice versa, which means that the $(u,v)$ domains of the triangles form an exact partition of $[0,1]^2$. Thus the triangle mesh is `water-tight', meaning that no ray intersection should be lost by rays passing in between triangles.

\subsection{Image reconstruction}
The ray tracing produces an irradiance distribution in the form of an image matrix $\mathcal{I} \in \mathbb{R}^{n_x \times n_y}_{\ge 0}$, where the elements correspond to a grid of rectangles called pixels that partition the detector screen positioned at $z=z_\text{screen} > \max_{i,j} P_{i,j}^z$. The screen resolution $(n_x,n_y)$ and the screen radii $(R_x,R_y)$ together yield the pixel size
\begin{equation}
    (w_x,w_y) = \left(\frac{2R_x}{n_x},\frac{2R_y}{n_y}\right).
\end{equation}
For reasons explained later in this section, sometimes a few `ghost pixels' are added, so the effective screen radii are
\begin{equation}
    R_x^* := R_x + \frac{\nu_x - 1}{2}w_x, \quad
    R_y^* := R_y + \frac{\nu_y - 1}{2}w_y,
\end{equation}
and the effective screen resolution is $(n_x + \nu_x -1, n_y + \nu_y - 1)$ where $\nu_x$ and $\nu_y$ are odd positive integers whose meaning will become clear later in this section.

Producing the irradiance distribution from the rays that intersect the detector screen is called image reconstruction \cite[sec. 7.8]{pharr2016physically}. The way that a ray contributes to a pixel with indices $i,j$ is governed by a reconstruction filter
\begin{equation}
    F_{i,j} : [-R_x,R_x] \times [-R_y,R_y] \rightarrow \mathbb{R}_{\ge 0},
\end{equation}
yielding for the irradiance distribution
\begin{equation}
    \mathcal{I}_{i,j} = \sum_{k=1}^N \omega_k F_{i,j}(\mathbf{x}_k),
\end{equation}
for a set of ray intersections $\{\mathbf{x}_k\}_{k=1}^N$ with corresponding final ray weights $\{\omega_k\}_{k=1}^N$. The ray weights are initialized at the sampling of the ray at the source. They are slightly modified by the lens boundary interactions as a small portion of the light is reflected rather than refracted. The amount by which the ray weights are modified is governed by the Fresnel equations \cite[sec. 2.7.1]{Fowles1975}. In our implementation, the Fresnel equations are approximated by Schlick's approximation \cite[eq. 24]{Schlick1994}. In the current implementation, all ray weights are initialized equally. The precise value does not matter since the relationship between the initial and final weights is linear. The loss function (section \ref{lossfunc}) compares scaled versions of the produced and target irradiance distribution.

In the simplest reconstruction case, the value of a pixel is given by the sum of the weights of the rays that intersect the detector screen at that pixel (called box reconstruction in \cite[sec. 7.8.1]{pharr2016physically}). In this case the reconstruction filter of pixel $i,j$ is simply the indicator function of the pixel $\left[(i-1)w_x,iw_x\right) \times \left[(j-1)w_y,jw_y\right)$.

\begin{figure}
    \centering
    \includegraphics[width = 0.7\textwidth]{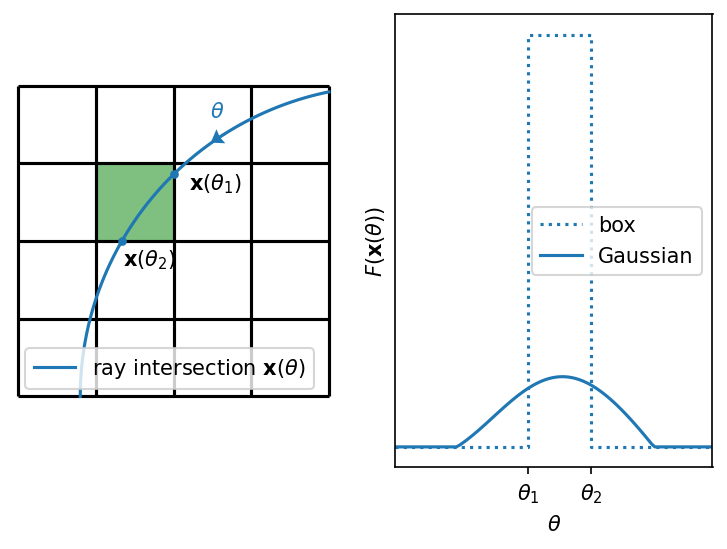}
    \caption{$\mathbf{x}(\theta)$ in the left plot shows the intersection location of a ray with the screen, dependent on a lens geometry parameter $\theta$. The right plot then shows the reconstruction filter value for the green pixel in the left plot dependent on $\theta$. In order to obtain a useful gradient of the pixel value with respect to $\theta$, a smooth reconstruction filter is needed.}
    \label{fig:reconstructiondiffb}
\end{figure}

To obtain a ray tracing implementation where the irradiance $\mathcal{I}$ is differentiable with respect to geometry parameters of the lens, say, the parameter $\theta$, the irradiance distribution must vary smoothly with this parameter. The dependency on this parameter is carried from the lens to the screen by the rays through the screen intersections $\mathbf{x}_k = \mathbf{x}_k(\theta)$. Thus to obtain a useful gradient $\frac{\partial \mathcal{I}}{\partial \theta}$ the filter function $F_{i,j}$ should be at least $C^1$, see Fig.~\ref{fig:reconstructiondiffb} which is achieved by introducing a filter function that spreads out the contribution of a ray over a kernel of pixels of size $(\nu_x,\nu_y)$ centered at the intersection location. For the conservation of light, we require that $\sum_{i,j}F_{i,j}(\mathbf{x}) \equiv 1$.

\begin{figure}
    \centering
    \includegraphics{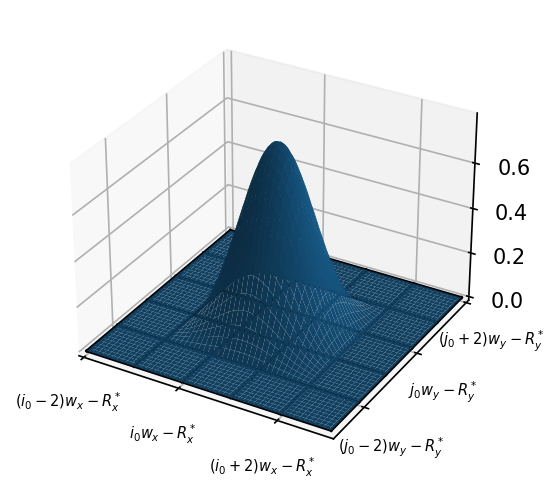}
    \caption{Gaussian reconstruction filter $F_{i_0,j_0}$ for $\alpha = 1$ and $(\nu_x,\nu_y)=(3,3)$.}
    \label{fig:recfilter3d}
\end{figure}

Therefore, the Gaussian reconstruction function is introduced, based on the identically named one described in \cite[sec. 7.8.1]{pharr2016physically}. This filter function is based on the product 
\begin{equation}
    \tilde{F}_{i,j}(x,y;\alpha,\nu_x,\nu_y) := f_{i}^x(x;\alpha,\nu_x)f_{j}^y(y;\alpha,\nu_y),
\end{equation}
where
\begin{equation}
    f_{i_0}^x(x;\alpha,\nu_x) = 
    \begin{cases}
        e^{-\alpha\left(x-c^x_{i_0}\right)^2} - e^{-\alpha\left(\frac{\nu_x w_x}{2}\right)^2} &\text{ if } \lvert x-c^x_i\rvert < \frac{\nu_x w_x}{2},\\
        0 & \text{otherwise.}
    \end{cases} \label{eq:filter1dim}
\end{equation}
The centers of the pixels are given by
\begin{equation}
    (c_i^x,c_j^y) := \left(\left(i + \textstyle\frac{1}{2}\right)w_x - R_x, \left(j + \textstyle\frac{1}{2}\right)w_y - R_y\right).
\end{equation}
Note that the support of $\tilde{F}_{i,j}$ is of size $\nu_xw_x$ by $\nu_yw_y$, the size of the kernel on the detector screen. The normalized reconstruction filter is then given by
\begin{equation}
    F_{i,j}(x,y;\alpha,\nu_x,\nu_y) = \frac{\tilde{F}_{i,j}(x,y;\alpha,\nu_x,\nu_y)}{\sum_{i',j'}\tilde{F}_{i',j'}(x,y;\alpha,\nu_x,\nu_y)}.
\end{equation}
The function $F_{i,j}$ is plotted in Fig.~\ref{fig:recfilter3d}. Note that the function is not differentiable at the boundary of its support, but this yields no problems in the optimization.

\begin{figure}
    \centering
    \includegraphics[width = \textwidth]{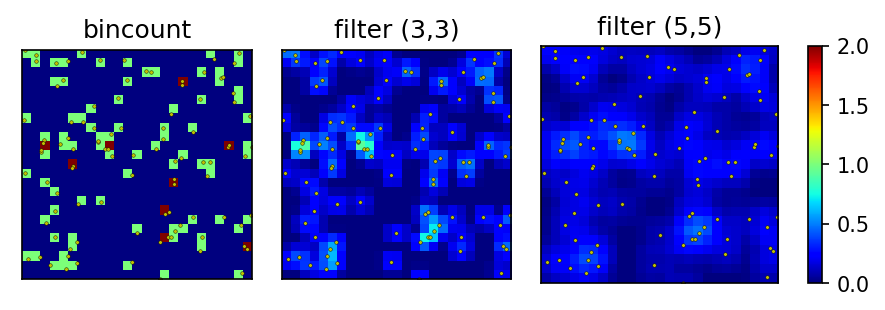}
    \caption{Image reconstruction based on a small set of ray-screen intersections, for bincount and various reconstruction filter sizes and $\alpha = 1$.}
    \label{fig:imagereconstr}
\end{figure}

Gaussian image reconstruction is shown in Fig.~\ref{fig:imagereconstr} for various values of $\nu_x = \nu_y$. There is a trade-off here since the larger $\nu_x$, and $\nu_y$ are the blurrier the resulting image is, and the larger the computational graph becomes, but also the larger the section of the image is that is aware of a particular ray which yields more informative gradients. 

Up to this point, this section has discussed the ray tracing part of the pipeline, the next subsections will discuss the role of the neural network and the optimization.

\subsection{Multi-layer perceptron as optimization accelerator} \label{subsec:nnarchitectures}
\begin{figure}
    \centering
    \includegraphics[width = 0.6\textwidth]{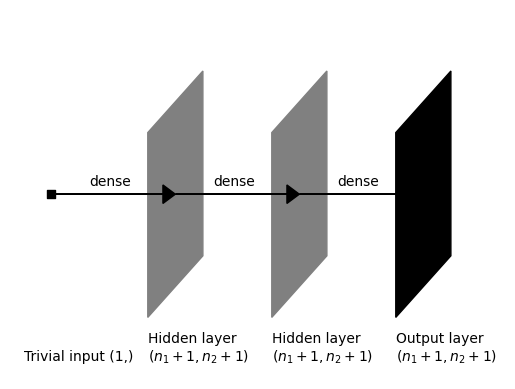}
    \caption{The dense multi-layer perceptron architecture based on the size of the control net $(n_1+1)\times(n_2+1)$.}
    \label{fig:densenn}
\end{figure}

Several neural network architectures are considered, all with a trivial input of 1, meaning that the neural networks will not, strictly speaking, be used to approximate a function since the considered domain is trivial. Non-trivial network inputs of system parameters like the source location will probably be part of follow-up research.

In this configuration, the neural network can be considered a transformation of the space over which is optimized: from the space of trainable neural network parameters to the space of control point $z$-coordinate values.
The goal of choosing the network architecture is that optimizing the trainable neural network parameters of this architecture yields better training behavior than optimizing the control point z-coordinate values directly. The used networks are multi-layer perceptions (MLPs), feed-forward networks consisting of several layers of neurons, as depicted in Fig.~\ref{fig:densenn}. The considered architectures are:
\begin{enumerate}
    \item No network at all. \label{item:nonetcase}
    \item A sparse MLP where the sparsity structure is informed by the overlap of the B-spline basis function supports on the knot spans. In other words: this architecture aims to precisely let those control points 'communicate' within the network that share influence on some knot span product on the B-spline surface, yielding a layer with the same connectivity as a convolutional layer with kernel size $(2p+1,2q+1)$. However, each connection has its own weight and each kernel its own bias, instead of only having a weight per element of the convolution kernel and one single bias for all kernels.
    \item Larger fully connected architectures are also considered, with 3 layers of control net size. Note that two consecutive such layers yield many weight parameters: $n^4$ for a square control net with `side length' $n$.
\end{enumerate}
The activation function used for all neurons is the hyperbolic tangent, which is motivated below.

\subsubsection{Control point freedom} \label{subsec:controlpointfreedom}
Control over the range of values that can be assumed by the control point $z$-coordinates is essential to make sure that the systems stays physical (as mentioned in Section~\ref{subsubsec:lensconstraints}), but also to be able to take into account restrictions imposed on the lens as part of mechanical construction in a real-world application. Note that the restriction $P_{i,j}^z > z_\text{in}$ for the control points being above the lens entrance surface is not critical for a collimated ray bundle simulation since, the entrance surface can be moved arbitrarily to the $-z$ direction without affecting the ray tracing. 

Since the final activation function $\tanh$ has finite range $(-1,1)$, this can easily be mapped to a desired interval $(z_{\min},z_{\max})$:
\begin{equation}
    y_{i,j} \mapsto z_{\min} + \textstyle\frac{1}{2} (y_{i,j} + 1)(z_{\max} - z_{\min}), \label{eq:outputcorrection}
\end{equation}
which can even vary per control point if desired. Here $y_{i,j}$ denotes an element of the total output $Y$ of the network. %\footnote{$Y$ can be considered a vector as in \ref{fig:NNarchitectures} or as a matrix in the formulas used here. This does not matter, as long as a consistent translation between the two is used.}
The above can also be used as an offset from certain fixed values:
\begin{equation}
    y_{i,j} \mapsto f\left(P^x_{i,j},P^y_{i,j}\right) + z_{\min} + \textstyle\frac{1}{2} (y_{i,j} + 1)(z_{\max} - z_{\min}). \label{eq:outputcorrectionwfunc}
\end{equation}
The resulting B-spline surface approximates the surface given by $f(x,y) + \textstyle\frac{1}{2}(z_{\max}+z_{\min})$ if $Y \approx 0$ can be used to optimize a lens that is globally at least approximately convex/concave. The choice of the hyperbolic tangent activation function accommodates this: since this activation function is smooth around its fixed point $0$ when initializing the weights and biases of the network close to $0$, there is no cumulative value-increasing effect in a forward pass through the network so that indeed $Y\approx 0$ in this case.

For comparability, in the case without a network, the optimization is not performed directly on the control point $z$-coordinates. Instead, for each control point, a new variable for optimization is created, which is passed through the activation function and the correction as in Eq.~\ref{eq:outputcorrection} or \ref{eq:outputcorrectionwfunc} before being assigned to the control point.

\subsection{The optimization}

\label{lossfunc}
The lens is optimized such that the irradiance distribution $\mathcal{I}$ projected by the lens approximates a reference image $\mathcal{I}_\text{ref}$, where $\mathcal{I} ,\mathcal{I}_\text{ref}\in \mathbb{R}^{n_x\times n_y}_{\ge 0 }$. The loss function used to calculate the difference between the two uses the normalized matrices: 
\begin{equation}
    \widehat{\mathcal{I}} = \frac{\mathcal{I}}{\sum_{i,j}^{n_x,n_y} \mathcal{I}_{i,j}} 
    \quad
    \mathrm{and}
    \quad
    \widehat{\mathcal{I}}_\text{ref} = \frac{\mathcal{I}_\text{ref}}{\sum_{i,j}^{n_x,n_y} \mathcal{I}_{\mathrm{ref},i,j}}.
\end{equation}

\noindent
The loss function is given by
\begin{equation}
    \mathcal{L}(\mathcal{I};\mathcal{I}_\text{ref}) = \frac{1}{\sqrt{n_x n_y}}
   \left\| \widehat{\mathcal{I}}-\widehat{\mathcal{I}}_\text{ref} \right\|_F 
    \label{eq:pipelineLoss},
\end{equation}
where $\| \cdot \|_F$ is the Frobenius or matrix norm, which is calculated as follows:
\begin{equation}
    \| \mathcal{A} \|_F = \sqrt{\sum_{i}^{n_x}\sum_{j}^{n_y} \lvert a_{i,j} \rvert ^2}.
\end{equation}
Fig.~\ref{fig:optimizationloop} shows the conventional stopping criterion of the loss value being smaller than some $\varepsilon > 0$, but in our experiments, we use a fixed number of iterations.

The neural network parameters (weights and biases) are updated using the Adam optimizer \citep{Kingma2014}  by back-propagation of the loss to these parameters.

\section{Results} \label{sec:results}

Several results produced with the optimization pipeline discussed in the previous sections are displayed and discussed in this section. The implementation mainly uses PyTorch, a Python wrapper of Torch \citep{Collobert2002Torch}.

None of the optimizations performed for this section took more than a few hours to complete, on a \verb|HP ZBook Power G7 Mobile Workstation| with a \verb|NVIDIA Quadro T1000 with Max-Q Design| GPU.

Most of the results have been validated with \emph{LightTools} \citep{LightTools}, an established ray tracing software package in the optics community. Lens designs were imported to LightTools as a point cloud, then interpolated to obtain a continuous surface, and all simulations were conducted using $10^6$ rays.

Units of length are mostly unspecified since the obtained irradiance distributions are invariant under uniform scaling of the optical system. This invariance to scaling is reasonable as long as the lens details are orders of magnitude larger than the wavelength of the incident light such that diffraction effects do not play a role. Furthermore, the irradiance distributions are directly proportional to the scaling of all ray weights and thus the source power, so the source and screen power also need no unit specification. Note that relative changes have a non-trivial effect, like changes to the power proportion between sources or the distance proportions of the optical system.

\newpage
\subsection{Irradiance derivatives with respect to a control point}
\begin{figure}
    \centering
    \includegraphics[width=0.75\textwidth]{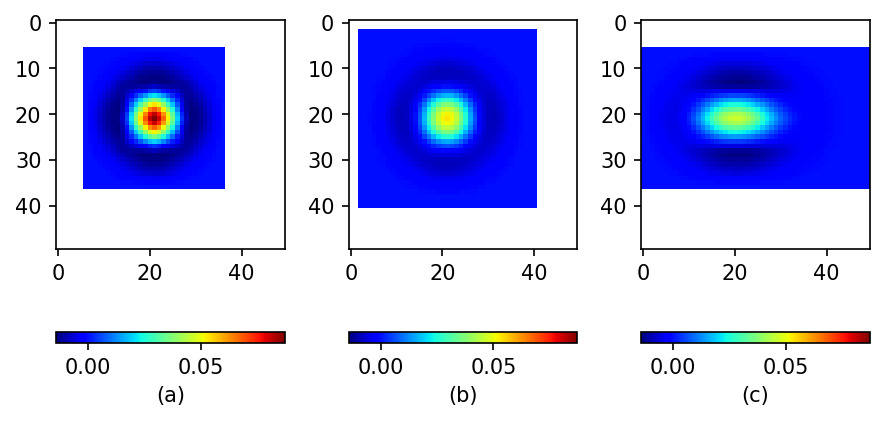}
    \caption{Gradients of an irradiance distribution of a collimated ray bundle through a flat lens (parallel sides), with respect to the $z$-coordinate of one control point. The zeros are masked with white to show the extend of the influence of the control point. These irradiation distributions differ by: (a): degrees $(3,3)$, reconstruction filter size $(3,3)$, (b): degrees $(3,3)$, reconstruction filter size $(11,11)$, (c): degrees $(5,3)$, reconstruction filter size $(3,3)$.}
    \label{fig:controlpointderivs}
\end{figure}

\begin{figure}
    \centering
    \includegraphics[width=0.9\textwidth]{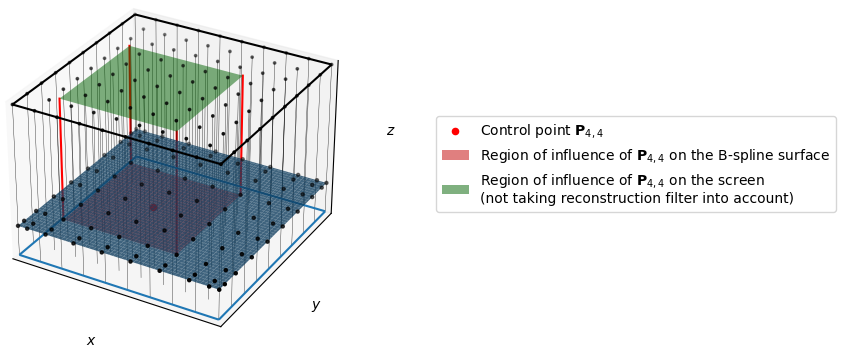}
    \caption{Demonstration of how one control point influences the irradiance distribution in the case of a flat lens with B-spline degrees $(3,3)$ and a collimated ray bundle source.}
    \label{fig:renderderivsexplanation}
\end{figure}
This section gives a simple first look at the capabilities of the implemented differentiable ray tracer: computing the derivative of an irradiance distribution with respect to a single control point. Obtaining this data is inefficient in the current PyTorch implementation as a forward mode automatic differentiation pass is required, which is not currently (entirely) supported by PyTorch. Therefore these derivatives are computed with pixel-wise back-propagation.

Fig.~\ref{fig:controlpointderivs} shows the derivative of an irradiance distribution produced by a collimated ray bundel through a flat lens for various B-spline degrees and reconstruction filter sizes, and Fig.~\ref{fig:renderderivsexplanation} shows what one of these systems looks like. The overall `mountain with a surrounding valley' structure can be understood as follows: as one of the control points rises, it creates a local convexity in the otherwise flat surface. This convexity has a focusing effect, redirecting light from the negative valley region toward the positive mountain region.

Noteworthy of these irradiance derivatives is also their total sum: (a) $\SI{-1.8161e-08}{}$, (b) $\SI{3.4459e-08}{}$, (c) $\SI{9.7095e-05}{}$. These small numbers with respect to the total irradiance of about $93$ and therefore indicate conservation of light; as the control point moves out of the flat configuration, at first, the total amount of power received by the screen will not change much. This is expected from cases (a) and (b), where the control point does not affect rays that reach the screen on the boundary pixels. However, in all cases, all rays intersect the lens at right angles. Around $\theta = 0$, the slope of Schlick's approximation is very shallow, indicating a small decrease in refraction in favor of reflection.

\subsection{Sensitivity of the optimization to initial state and neural network architecture} \label{subsec:results_nnsensitivity}
\begin{figure}
    \centering
    \includegraphics[width=0.4\textwidth]{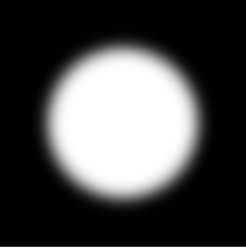}
    \caption{The circular tophat target illumination.}
    \label{fig:circtophat}
\end{figure}

\begin{figure}
    \centering
    \includegraphics[width=0.4\textwidth]{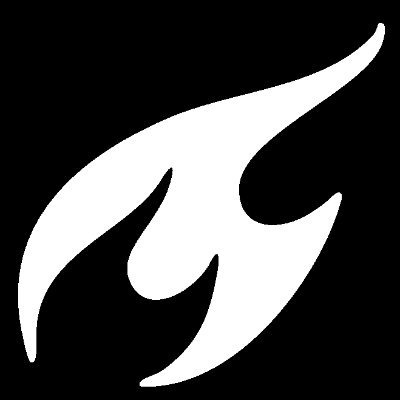}
    \caption{The TU Delft flame target illumination.}
    \label{fig:TUDflameinv}
\end{figure}

\begin{figure}
    \centering
    \includegraphics[width = 0.4\textwidth]{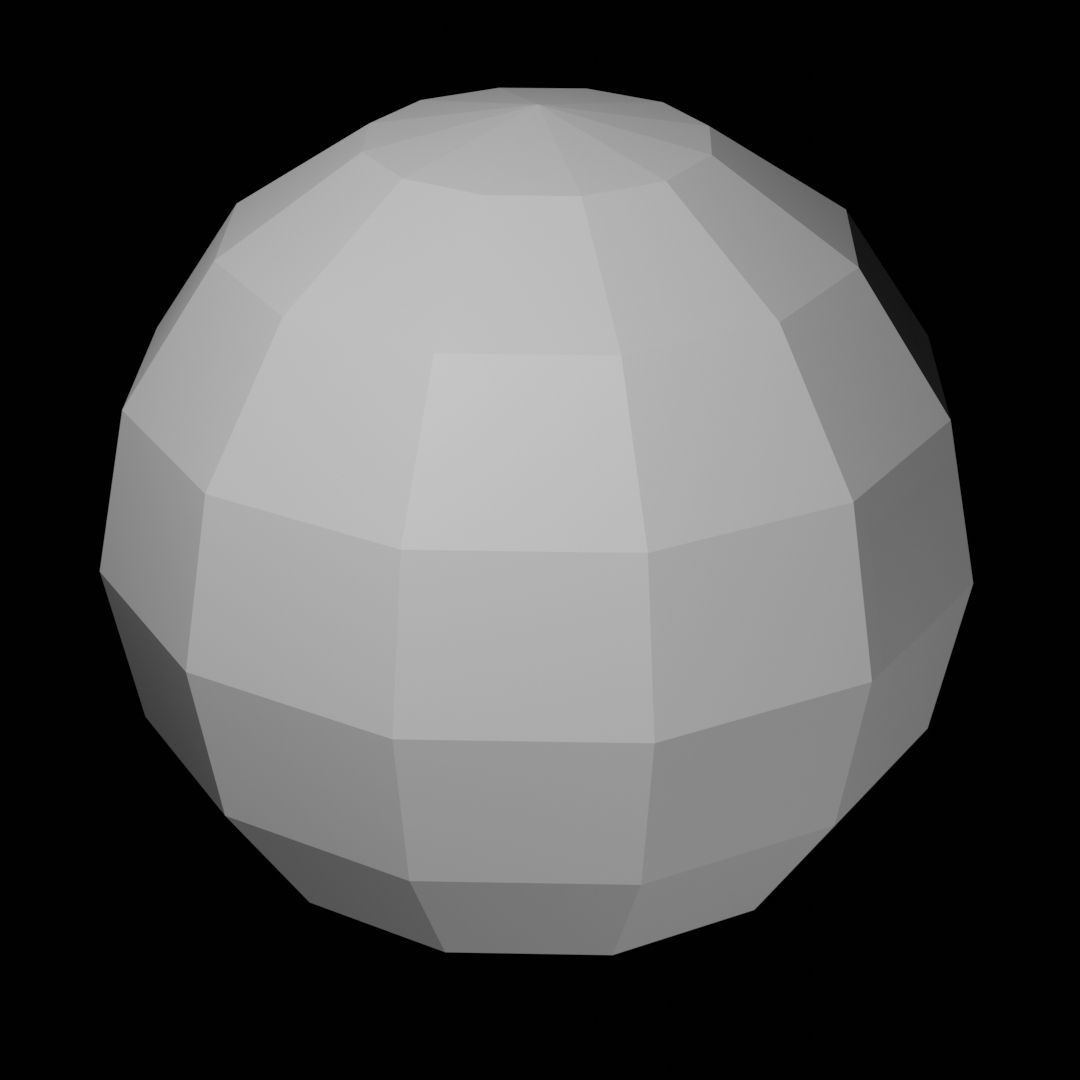}
    \caption{The faceted ball target illumination.}
    \label{fig:facball}
\end{figure}

As with almost any iterative optimization procedure, choosing a reasonable initial guess of the solution is crucial for reaching a good local/global minimum. For training neural networks, this comes down to how the network weights and biases are initiated. In this section, we look at three target illuminations: the circular top hat distribution (Fig.~\ref{fig:circtophat}), the TU Delft logo (Fig.~\ref{fig:TUDflameinv}), and an image of a faceted ball (Fig.~\ref{fig:facball}). For some experiments, black padding or Gaussian blurring is applied to these images. We design lenses to produce these distributions from a collimated ray bundle, given various neural network architectures (introduced in section \ref{subsec:nnarchitectures}) and parameter initializations.

\begin{figure}[h]
    \centering
    \includegraphics[width=0.7\textwidth]{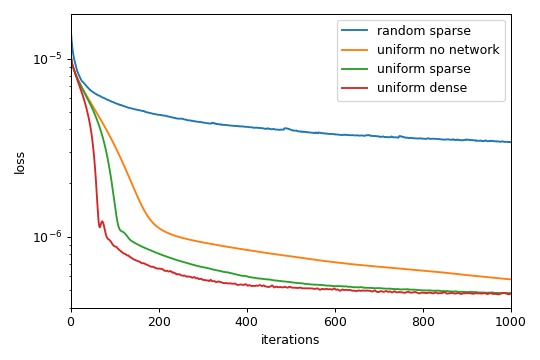}
    \caption{Loss progress over the iterations for various pipeline-setups for forming a tophat distribution from a collimated ray bundle.}
    \label{fig:tophatloss}
\end{figure}

\begin{figure}
    \centering
    \includegraphics[width=0.9\textwidth]{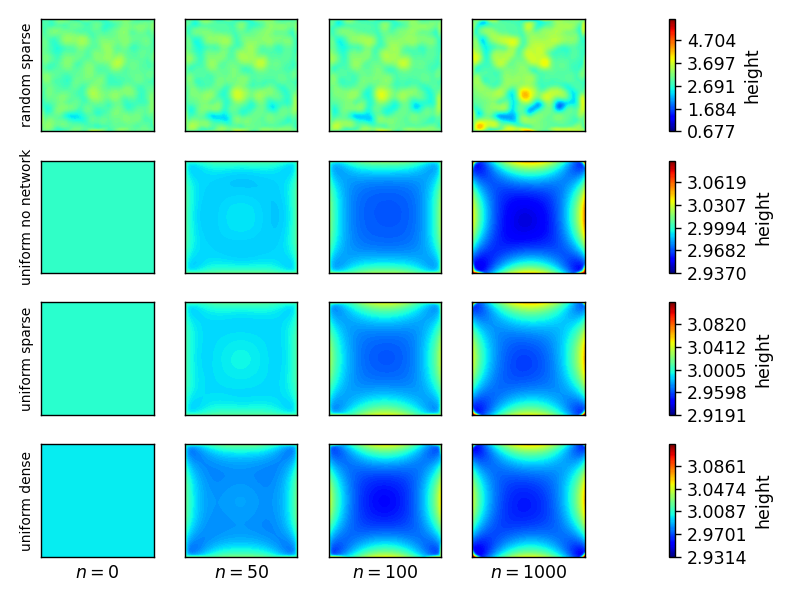}
    \caption{The lens height field after initialization ($n=0$), and $n=50,100$ and $1000$ iterations respectively, for different network architectures (Section~\ref{subsec:nnarchitectures}) and network parameter initializations (Section~\ref{subsec:circtophat}).}
    \label{fig:lensshapes}
\end{figure}

\begin{figure}[p]
    \centering
    \includegraphics[width=0.9\textwidth]{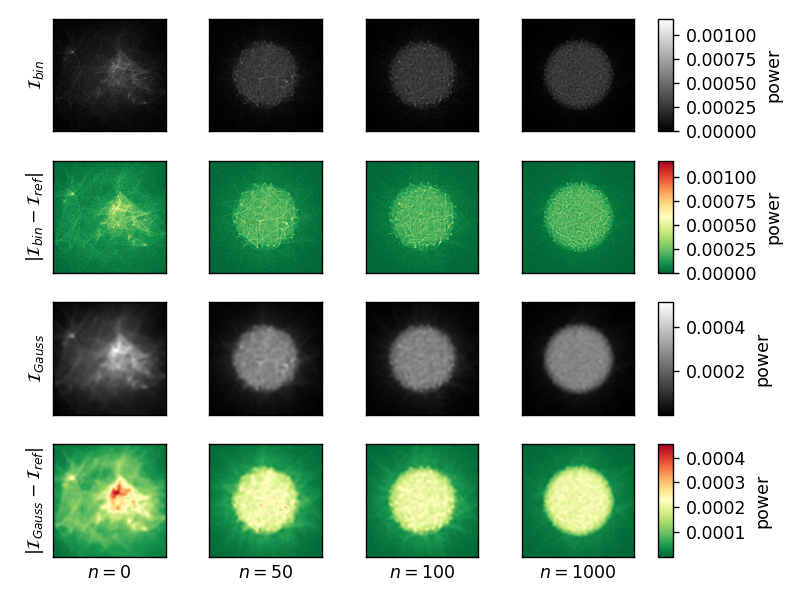}
    \caption{Irradiance distributions and pixel-wise errors in the optimization progress of a random lens with a sparse network towards a circular tophat illumination.}
    \label{fig:tophatrandomsparse}
\end{figure}

\begin{figure}[p]
    \centering
    \includegraphics[width=0.9\textwidth]{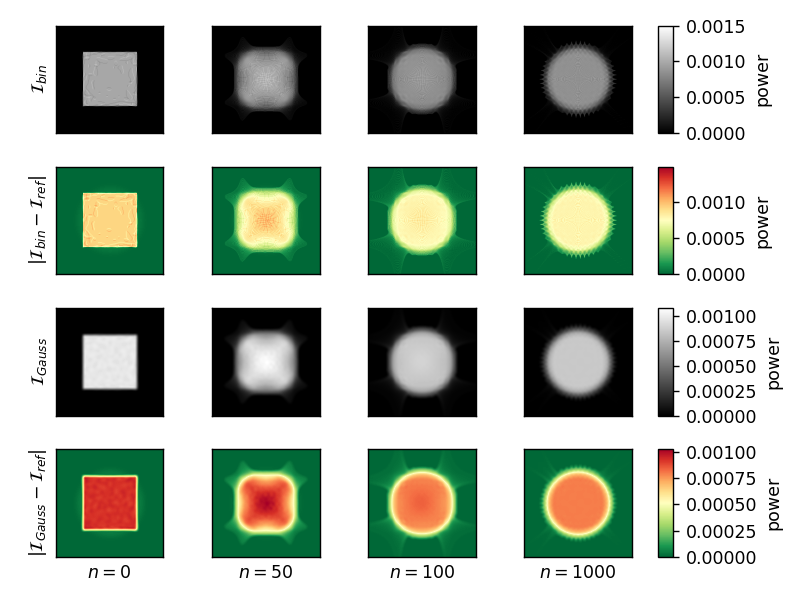}
    \caption{Irradiance distributions and pixel-wise errors in the optimization progress of a flat lens with a sparse network towards a circular tophat illumination.}
    \label{fig:tophatunifsparse}
\end{figure}

\begin{figure}[p]
    \centering
    \includegraphics[width=0.9\textwidth]{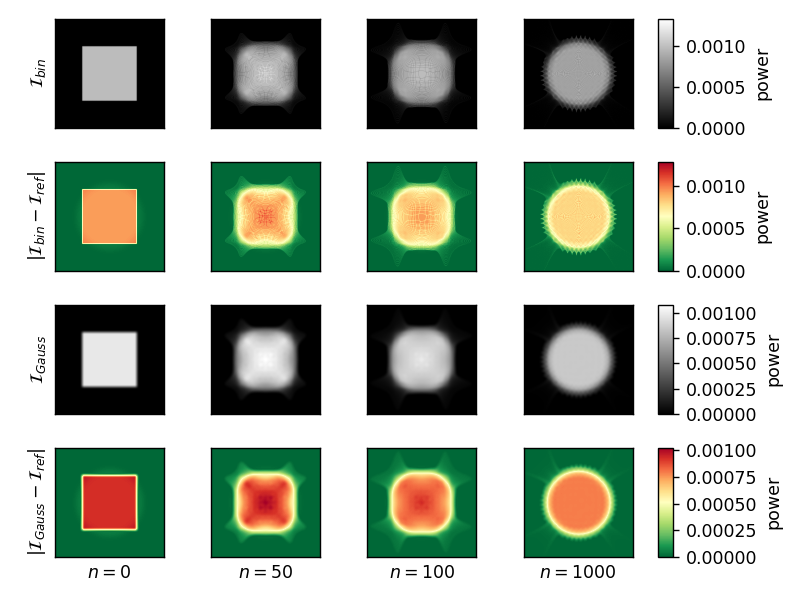}
    \caption{Irradiance distributions and pixel-wise errors in the optimization progress of a flat lens without a network towards a circular tophat illumination.}
    \label{fig:tophatunifnonet}
\end{figure}

\begin{figure}[p]
    \centering
    \includegraphics[width=0.9\textwidth]{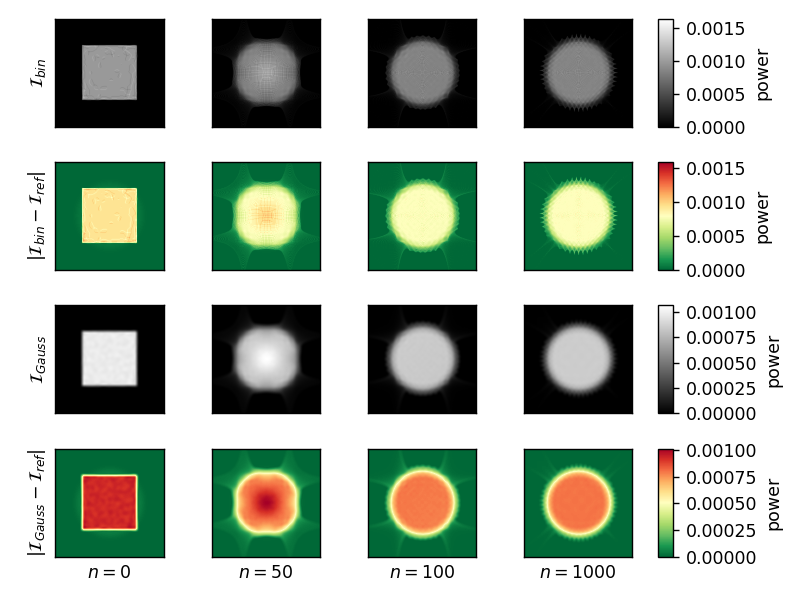}
    \caption{Irradiance distributions and pixel-wise errors in the optimization progress of a flat lens with a dense network towards a circular tophat illumination.}
    \label{fig:tophatunifdense}
\end{figure}

\paragraph{Circular top hat distribution from collimated ray bundle} \label{subsec:circtophat}
Fig.~\ref{fig:tophatloss} shows the progress of the loss over 1000 iterations, with each iteration taking $2.5$ seconds, for various neural network architectures and parameter initialization combinations. For the other parameters in these simulations, see the supplementary information. For a few moments during the training, the resulting freeform surfaces and irradiance distributions are shown in Figs.~\ref{fig:lensshapes}, \ref{fig:tophatrandomsparse}, \ref{fig:tophatunifsparse}, \ref{fig:tophatunifnonet} and \ref{fig:tophatunifdense}. Uniform here means that the initial trainable parameter values are sampled from a small interval: $U\left(\left[-10^{-4},10^{-4}\right]\right)$, except for the no-network case; this is initialized with all zeros.

The first notable difference is between the random and uniformly initialized sparse neural networks. The uniformly initialized neural network performs much better, and no network performs better. This is probably because the uniformly initialized cases converge to a better (local) minimum than the randomly initialized case. Of course, it could happen that the random initialization lands in a very favorable spot in the design landscape, but intuitively this seems very unlikely.

Another property of the uniformly initialized cases is their preservation of symmetry in these setups. As Fig.~\ref{fig:lensshapes} shows, this leads to much simpler lenses, which are probably much less sensitive to manufacturing errors due to their relative lack of small detail. What is interesting to note here is that if the sparse network is initialized with all parameters set to $0$, then its optimization is identical to the no-network case, as only the biases in the last layer achieve non-zero gradients.

No rigorous investigation has been conducted to the extent that this behavior of increased convergence speed carries over to other target distributions and system configurations and what the optimal hyper-parameters are. A thorough investigation of the hyper-parameter space that defines a family of network architectures could reveal where in the increase of the architecture complexity, diminishing returns for optimizing these lenses arises. However, based on these initial findings the fully connected network is used for all the following optimizations in the results.

\clearpage
\paragraph{TU flame and faceted ball from collimated ray bundle}
In what follows, we consider complex target distributions: the TU Delft flame (for a complex shape) and a faceted ball (for a target with various brightness levels). Here we still use the collimated ray bundle illumination, but lenses are now optimized for various magnifications; see Table.~\ref{tab:magndata}. These magnifications are defined as the scaling of the screen size with respect to the smallest screen size $(0.64,0.64)$.  The other parameters of these optimizations are shown in the supplementary information. All these iterations took about $4$ seconds each.

\begin{figure}[h]
    \centering
    \includegraphics[width=0.7\textwidth]{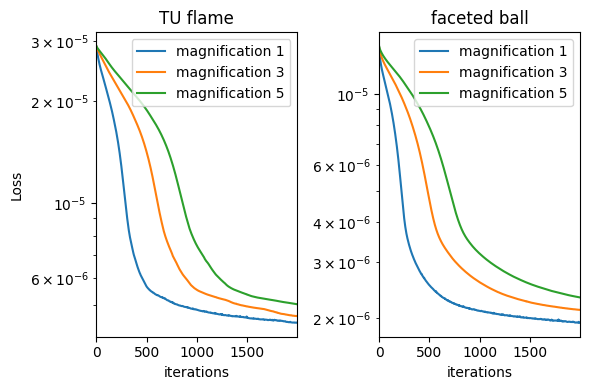}
    \caption{Loss progress for the various magnifications and target distributions.}
    \label{fig:ManufacturingLoss}
\end{figure}

\begin{figure}[p]
    \centering
    \includegraphics[width=\textwidth]{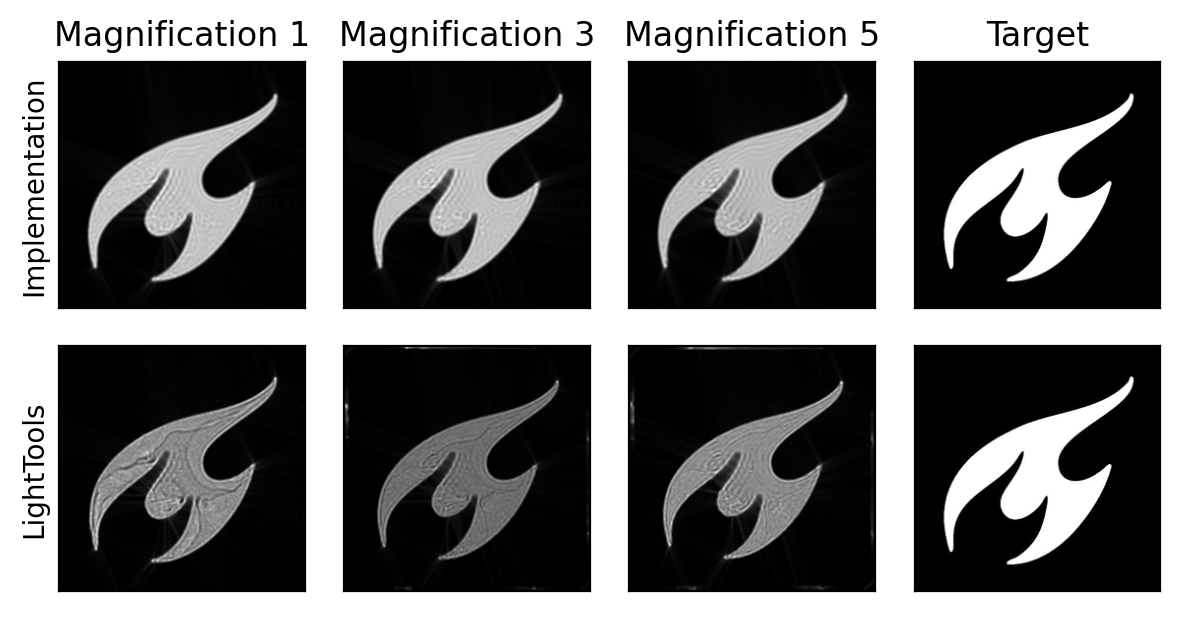}
    \caption{Implementation and LightTools irradiance distributions of the TU flame target from the final lens design of the optimization.}
    \label{fig:flamerenders}
\end{figure}

\begin{figure}[p]
    \centering
    \includegraphics[width=\textwidth]{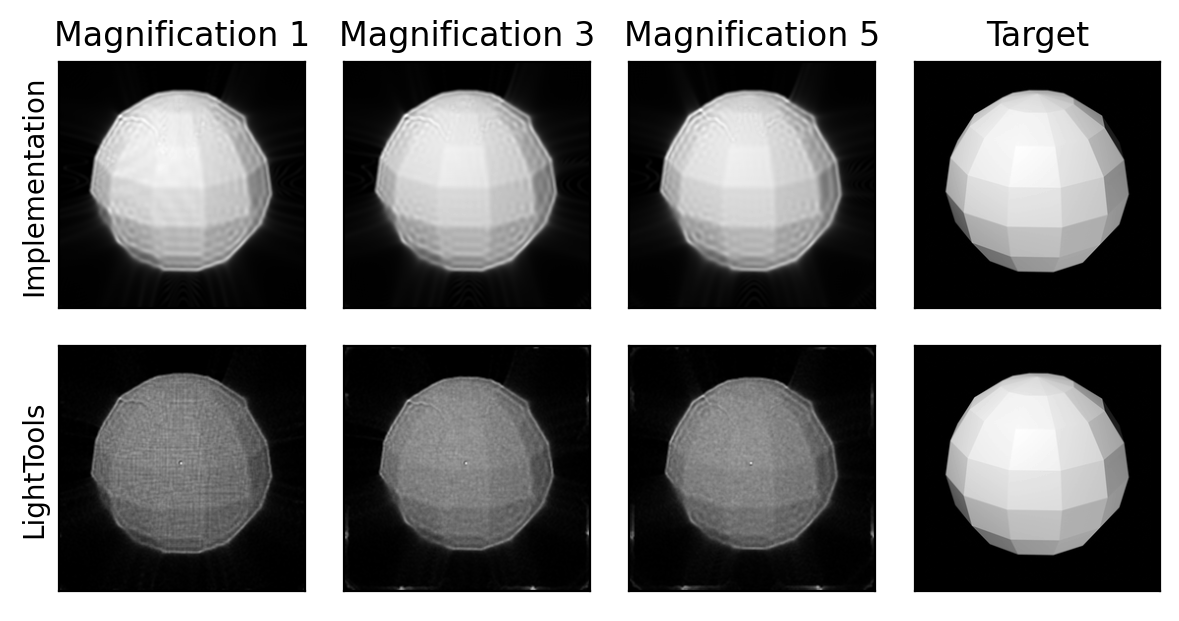}
    \caption{Implementation and LightTools irradiance distributions of the faceted ball target from the final lens design of the optimization.}
    \label{fig:facetedballrenders}
\end{figure}

The final irradiance distributions and corresponding LightTools results are shown in Figs.~\ref{fig:flamerenders} and \ref{fig:facetedballrenders}, respectively. These figures show that the optimization pipeline can handle these more complex target illuminations well. The LightTools results predict some artifacts within the irradiance distribution, which the implemented ray tracer does not, especially in the TU flame magnification 1 case. By visual inspection, based on the LightTools results, one would probably rate these results in the exact opposite order than as indicated by the losses shown in Fig.~\ref{fig:ManufacturingLoss}.

\begin{figure}[p]
    \centering
    \includegraphics[width=\textwidth]{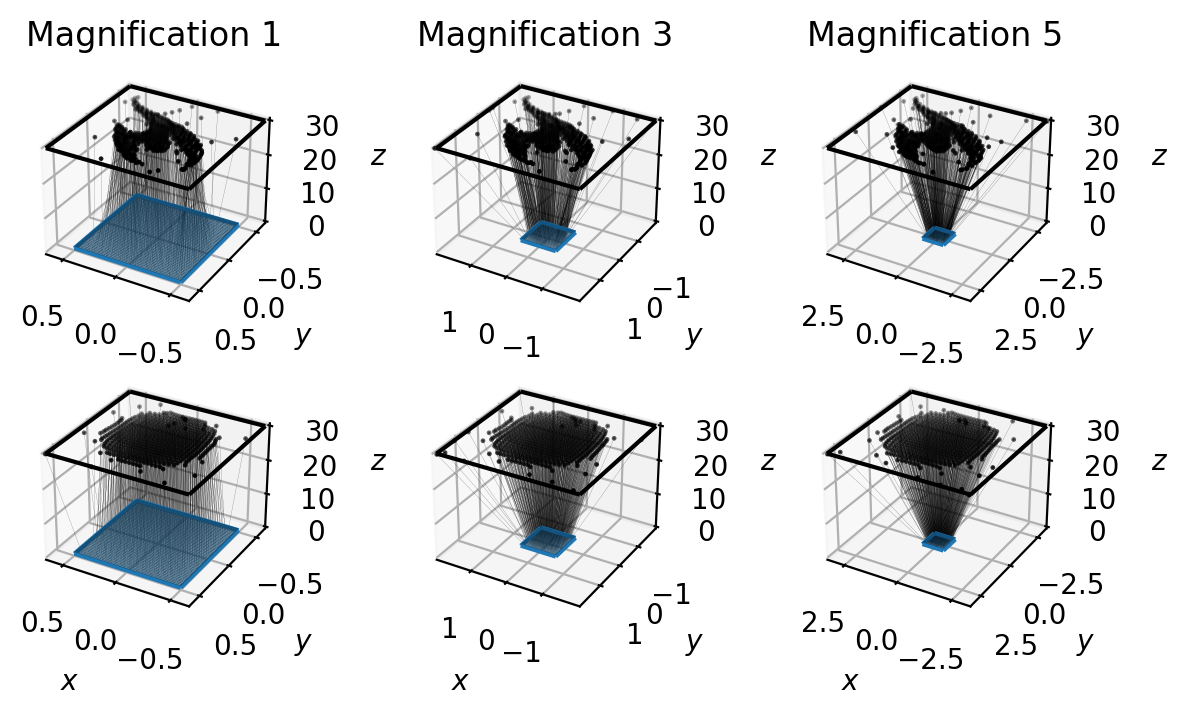}
    \caption{$25\times25$ traced rays through the final lens designs for the different magnifications and two target distributions.}
    \label{fig:ManufacturingRays}
\end{figure}

A potential explanation of the increase in loss with the magnification factor in Fig.~\ref{fig:ManufacturingLoss} is that the bigger the screen is: the rays require higher angles to reach the edges of the screen, which is apparent in the cases of magnification 3 and 5 Fig.~\ref{fig:ManufacturingRays}. This results in a larger sensitivity of the irradiance to the angle with which a ray leaves the screen. This in turn gives larger gradients of the irradiance with respect to the control points. Therefore the optimization takes larger steps in the neural network parameter space, possibly overshooting points that result in a lower loss.

For the magnification, $3$ and $5$, the irradiance distributions from LightTools show artifacts at the screen boundaries. A possible explanation for this is that the way the B-spline surfaces are transferred to LightTools is inaccurate at the surface boundaries.\footnote{Assuming only rays from the B-spline surface boundaries reach the screen boundary area.} This is because surface normals are inferred from fewer points on the B-spline surface at the boundary than in the middle of the surface by LightTools.

Furthermore, a significant amount of rays are lost during optimization because the target illuminations are black at the borders, so rays near the screen boundary will be forced off the screen by the optimization. Once rays are off the screen, they no longer contribute to the loss function. Once a ray misses the screen, the patch on the B-spline surface these rays originate from does not influence the irradiance and, thus, the loss function. 
However, this does not mean that this patch is idle for the rest of the optimization, as this patch can be in support of a basis function that corresponds to a control point that still affects rays that hit the screen. Therefore, the probability of getting idle lens patches with this setup decreases with the B-spline degrees since these determine the size of the support of the B-spline basis functions but might, in some cases, lead to oscillatory behavior, with rays alternating between hitting and missing the screen. 

\begin{figure}[p]
    \centering
    \includegraphics[width=\textwidth]{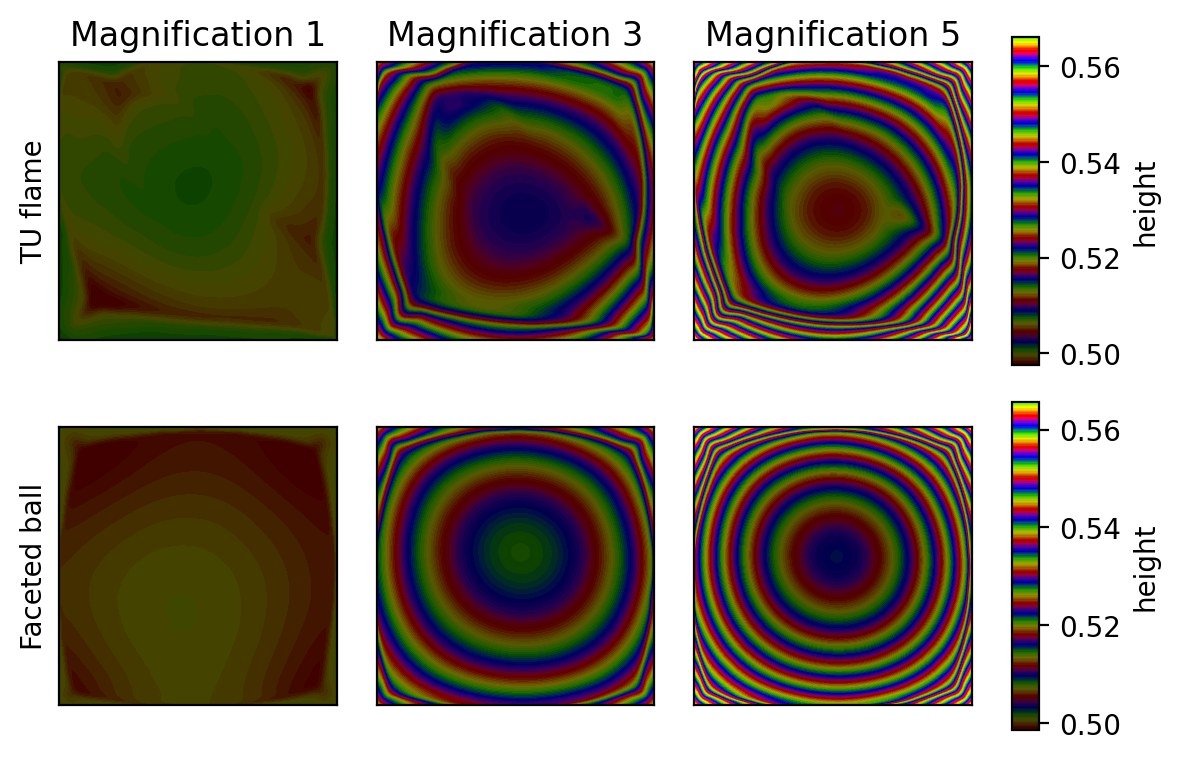}
    \caption{The lens designs for the different magnifications and two target distributions.}
    \label{fig:ManufacturingSurfaces}
\end{figure}

Fig. \ref{fig:ManufacturingSurfaces} shows the optimized B-spline lens surface height field. A densely varying color map is chosen since the deviations from a flat or smooth concave shape are quite subtle, which is due to the large lens exit angle sensitivity of the ray-screen intersections since the ratio lens size to screen size is large with respect to the ratio lens size to screen distance.

\begin{table}[h]
    \centering
    {
    \def\arraystretch{1.5}
        \begin{tabular}{c|c|c|c}
        \textbf{Magnification} & \textbf{screen size} & $f(x,y)$ & \textbf{starting shape type}\\
        \hline
        $1$ & $(0.64,0.64)$ & $\smolhalf$ & flat\\
        $3$ & $(1.92,1.92)$ & $\smolhalf + 8 - \sqrt{8^2-x^2-y^2}$ & concave\\
        $5$ & $(3.20,3.20)$ & $\smolhalf + 4 - \sqrt{4^2 - x^2 - y^2}$ & concave
        \end{tabular}
    }
    \caption{The screen size and control point offset function $f$ used per magnification in the TU flame and faceted ball optimizations (distances in centimeters).}
    \label{tab:magndata}
\end{table}

\clearpage

\clearpage
\subsection{Optimization with a point source and a grid of point sources}
\begin{figure}[p]
    \centering
    \includegraphics[width=\textwidth]{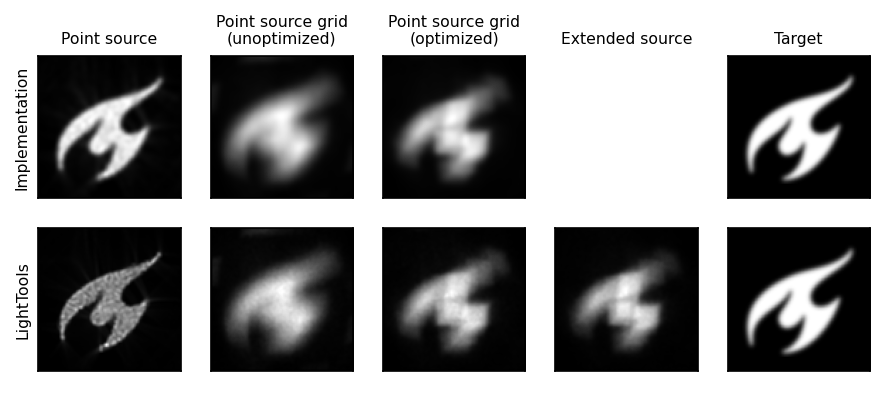}
    \caption{The final irradiation distribution of the lens optimizations with point sources and the corresponding LightTools verifications. The extended source is not implemented in our ray tracer, but is approximated by the point source grid.}
    \label{fig:point_source_renders}
\end{figure}

We now consider an optimization that uses the B-spline intersection algorithm. First, we design a lens with one point source at $(0,0,-5)$ with $5\times 10^5$ rays to again form the TU flame. Then after $\sim 200$ iterations, we change the source to an equispaced grid of $25\times 25$ point sources with $10^3$ rays each on $[-1,1]\times[-1,1]\times\{-5\}$, approximating a source of non-negligible size. The other (hyper-)~parameters of this optimization are shown in the supplementary information. Due to the additional B-spline intersection procedures, each iteration takes approximately $50$ seconds.
The resulting final irradiance distribution and LightTools verifications can be seen in Fig.~\ref{fig:point_source_renders}. The final irradiance distribution similar to the that obtained by LightTools, indicating that ray tracing with the implemented B-spline intersection algorithm works correctly. The irradiance are blurred due to the reconstruction filter.
The single-source point optimization performs well, although the illumination is less uniform than in the collimated ray bundle case (Figs.~\ref{fig:flamerenders} and \ref{fig:facetedballrenders}). 
The non-uniformity can be attributed to the gaussian reconstruction filter used during optimization, as it smoothes out the small uniformities.

\begin{figure}
    \centering
    \includegraphics{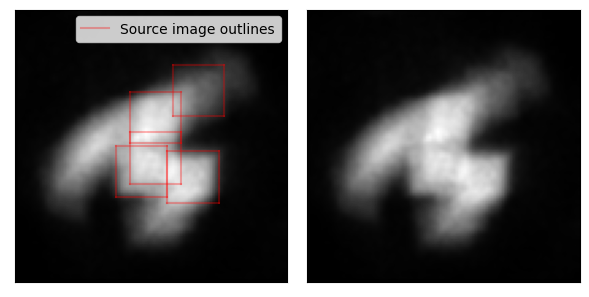}
    \caption{Indication of images of the source square in the irradiance distribution obtained by LightTools using the point source grid.}
    \label{fig:sourceimages}
\end{figure}

\begin{figure}[p]
    \centering
    \includegraphics[width= 0.6\textwidth]{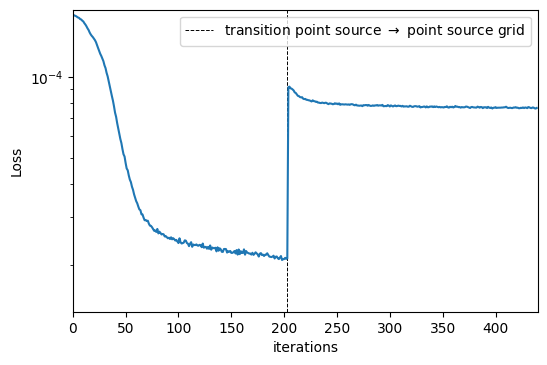}
    \caption{Loss over the iterations optimizing for the TU flame. The system is initiated with a point source, and after $\sim 200$ iterations the point source is replaced by an equispaced grid of $25\times 25$ point sources.}
    \label{fig:point_source_loss}
\end{figure}

As seen in Fig.~\ref{fig:point_source_renders} the irradiance distribution obtained with a grid of point sources accurately approximates the extended source illumination distribution quite well for the unoptimized case. 
Finding a lens design that redirects light from a source of non-negligible size into a desired irradiance distribution is a complex problem for which it is hard to indicate how good the optimal irradiance distribution can become.
The progress of the loss, as seen in Fig.~\ref{fig:point_source_loss}, shows that the optimization can still improve the loss, even after the transition to the grid of point sources. Interestingly, looking at Fig.~\ref{fig:point_source_renders} again, the optimization seems to adopt the coarse strategy of filling up the target distribution with images of the source square, as shown in Fig.~\ref{fig:sourceimages}. This strategy does hinder the possible quality of the final irradiance distribution as the image of the source on the target is larger than the fine details in the desired irradiance. Optimizing both the front and back surfaces of the freeform could resolve this issue, as this will cause the image of the source to change shape depending on where it ends up on the target screen.

\begin{figure}[p]
    \centering
    \includegraphics[width=0.75\textwidth]{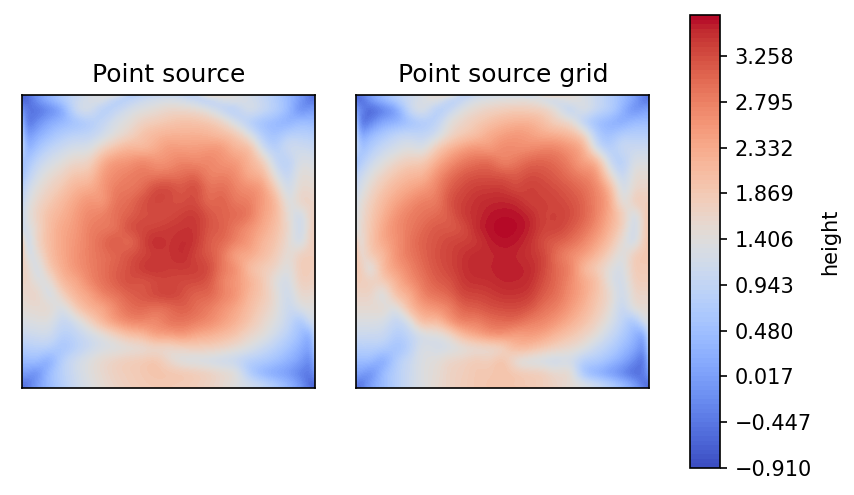}
    \caption{Height fields of the lenses optimized for the TU flame with point sources.}
    \label{fig:point_source_surfaces}
\end{figure}

\section{Conclusion} \label{sec:conclusion}

We demonstrated that non-sequential differentiable ray tracing is a viable tool for designing freeform lenses for collimated ray bundles, points, and extended sources.
Using a B-spline allows for the design of a continuous surface, which is desirable for manufacturing, and its control point allows for locally altering the irradiance distribution. For both cases, collimated and point source lens designs were found that could accurately project the desired irradiance distribution in both the differentiable ray tracer and in commercial software LightTools. Some artifacts still exist and resolving this issue will be a part of further research.

For the source with a finite extent, the optimizer improved upon the design obtained for a point source. However, the final irradiance distribution was made up of images of the source, which hinders the minimum that can be obtained as the source image is larger than the details in the desired irradiance distribution. This issue can be resolved by optimizing multiple surfaces simultaneously, as the image of the source on the target plane can then be optimized to vary with location.

Using a neural network to remap the optimization space provides an interesting way to increase the convergence speed of the optimization. However, further investigation is required to see whether this generally holds and what the effect is on other network architectures.

The developed ray tracing implementation is currently a proof of concept and needs to be optimized for speed. The B-spline intersection algorithm, in particular, adds roughly a factor of $10$ to the computation time. A significant speedup can be achieved here by leveraging efficient lower-level GPU programming languages, such as CUDA.

\section{Acknowledgements}
We acknowledge support by NWO-TTW Perspectief program (P15-36) ``Free-form scattering optics".
% \rem{For book entries in the bibliography `city' must be changed to `address' to avoid `???'}
\bibliography{bibliography}% common bib file

\begin{thebibliography}{}
\providecommand{\doi}[1]{\url{https://doi.org/#1}}
\bibcommenthead

\bibitem[\protect\citeauthoryear{Ansys}{Ansys}{2023}]{zemax}
Ansys. 2023.
\newblock Zemax.
\newblock https://www.zemax.com/.

\bibitem[\protect\citeauthoryear{Anthonissen, Romijn, ten Thije~Boonkkamp, and
  IJzerman}{Anthonissen et~al.}{2021}]{Anthonissen2021}
Anthonissen, M.J.H., L.B. Romijn, J.H.M. ten Thije~Boonkkamp, and W.L.
  IJzerman. 2021, 9.
\newblock Unified mathematical framework for a class of fundamental freeform
  optical systems.
\newblock {\em Optics Express\/}~29: 31650.
\newblock \doi{10.1364/oe.438920} .

\bibitem[\protect\citeauthoryear{Chen and Lin}{Chen and
  Lin}{2012}]{chen_second-order_2012}
Chen, Y.B. and P.D. Lin. 2012, August.
\newblock Second-order derivatives of optical path length of ray with respect
  to variable vector of source ray.
\newblock {\em Applied Optics\/}~{\em 51\/}(22): 5552.
\newblock \doi{10.1364/AO.51.005552} .

\bibitem[\protect\citeauthoryear{Cohen, Martin, Kirby, Lyche, and
  Riesenfeld}{Cohen et~al.}{2010}]{Cohen2010Iso}
Cohen, E., T.~Martin, R.M. Kirby, T.~Lyche, and R.F. Riesenfeld. 2010, 1.
\newblock Analysis-aware modeling: Understanding quality considerations in
  modeling for isogeometric analysis.
\newblock {\em Computer Methods in Applied Mechanics and Engineering\/}~199:
  334--356.
\newblock \doi{10.1016/j.cma.2009.09.010} .

\bibitem[\protect\citeauthoryear{Collobert, Bengio, and
  Mari{\'e}thoz}{Collobert et~al.}{2002}]{Collobert2002Torch}
Collobert, R., S.~Bengio, and J.~Mari{\'e}thoz 2002.
\newblock Torch: a modular machine learning software library.
\newblock Technical report, Idiap.

\bibitem[\protect\citeauthoryear{Falaggis, Rolland, Duerr, and Sohn}{Falaggis
  et~al.}{2022}]{Falaggis2022}
Falaggis, K., J.~Rolland, F.~Duerr, and A.~Sohn. 2022, 2.
\newblock Freeform optics: introduction.
\newblock {\em Optics Express\/}~30: 6450.
\newblock \doi{10.1364/oe.454788} .

\bibitem[\protect\citeauthoryear{Farin}{Farin}{2002}]{Farin2002}
Farin, G. 2002.
\newblock {\em Curves and Surfaces for CAGD\/} (5th ed.).
\newblock Burlington, Massachusetts: Morgan Kaufmann Publishers.

\bibitem[\protect\citeauthoryear{Feder}{Feder}{1968}]{feder_differentiation_1968}
Feder, D.P. 1968.
\newblock Differentiation of ray-tracing equations with respect to construction
  parameters of rotationally symmetric optics.
\newblock {\em JOSA\/}~{\em 58\/}(11): 1494--1505 .

\bibitem[\protect\citeauthoryear{Fowles}{Fowles}{1975}]{Fowles1975}
Fowles, G.R. 1975.
\newblock {\em Introduction to Modern Optics (2nd Edition)}.
\newblock Dover: Dover Publications.

\bibitem[\protect\citeauthoryear{Gasick and Qian}{Gasick and
  Qian}{2023}]{GASICK2023115839}
Gasick, J. and X.~Qian. 2023.
\newblock Isogeometric neural networks: A new deep learning approach for
  solving parameterized partial differential equations.
\newblock {\em Computer Methods in Applied Mechanics and Engineering\/}~405:
  115839.
\newblock \doi{https://doi.org/10.1016/j.cma.2022.115839} .

\bibitem[\protect\citeauthoryear{Givoli}{Givoli}{2021}]{givoli_tutorial_2021}
Givoli, D. 2021.
\newblock A tutorial on the adjoint method for inverse problems.
\newblock {\em Computer Methods in Applied Mechanics and Engineering\/}~380:
  113810.
\newblock \doi{10.1016/j.cma.2021.113810} .

\bibitem[\protect\citeauthoryear{Grant}{Grant}{2011}]{Grant2011}
Grant, B.G. 2011.
\newblock {\em Field guide to radiometry}.
\newblock Bellingham, Wash: SPIE.

\bibitem[\protect\citeauthoryear{John}{John}{2013}]{John2013}
John, R.K. 2013.
\newblock {\em Illumination Engineering: Design with Nonimaging Optics}.
\newblock Piscataway, NJ : Hoboken, New Jersey: John Wiley and Sons.

\bibitem[\protect\citeauthoryear{Karniadakis, Kevrekidis, Lu, Perdikaris, Wang,
  and Yang}{Karniadakis et~al.}{2021}]{Karniadakis2021}
Karniadakis, G.E., I.G. Kevrekidis, L.~Lu, P.~Perdikaris, S.~Wang, and L.~Yang.
  2021, 5.
\newblock Physics-informed machine learning.
\newblock {\em Nature Reviews Physics\/}~3: 422--440.
\newblock \doi{10.1038/s42254-021-00314-5} .

\bibitem[\protect\citeauthoryear{Kingma and Ba}{Kingma and
  Ba}{2014}]{Kingma2014}
Kingma, D.P. and J.~Ba. 2014.
\newblock Adam: A method for stochastic optimization.
\newblock Preprint at \url{http://arxiv.org/abs/1412.6980}.

\bibitem[\protect\citeauthoryear{Kronberg, Anthonissen, ten Thije~Boonkkamp,
  and IJzerman}{Kronberg et~al.}{2022}]{Kronberg2022}
Kronberg, V., M.~Anthonissen, J.~ten Thije~Boonkkamp, and W.~IJzerman. 2022.
\newblock Two-dimensional freeform reflector design with a scattering surface.
\newblock Preprint at \url{https://arxiv.org/abs/2211.03629}.

\bibitem[\protect\citeauthoryear{Lippman and Schmidt}{Lippman and
  Schmidt}{2020}]{lippman_prescribed_2020}
Lippman, D.H. and G.R. Schmidt. 2020.
\newblock Prescribed irradiance distributions with freeform gradient-index
  optics.
\newblock {\em Opt. Express\/}~28: 29132--29147.
\newblock \doi{10.1364/OE.404456} .

\bibitem[\protect\citeauthoryear{ltioptics}{ltioptics}{2023}]{photopia}
ltioptics. 2023.
\newblock Photopia.
\newblock https://www.ltioptics.com/en/optical-design-software-photopia.html.

\bibitem[\protect\citeauthoryear{Mohedano, Chaves, and Hernández}{Mohedano
  et~al.}{2016}]{Mohedano2016}
Mohedano, R., J.~Chaves, and M.~Hernández. 2016.
\newblock Free-form illumination optics.
\newblock {\em Advanced Optical Technologies\/}~5: 177--186.
\newblock \doi{10.1515/aot-2016-0006} .

\bibitem[\protect\citeauthoryear{M\"oller, Toshniwal, and van Ruiten}{M\"oller
  et~al.}{2021}]{Moller2021PIML}
M\"oller, M., D.~Toshniwal, and F.~van Ruiten. 2021.
\newblock Physics-informed machine learning embedded into isogeometric
  analysis, {\em Mathematics: Key enabling technology for scientific machine
  learning},  57--59. Amsterdam: Platform Wiskunde.

\bibitem[\protect\citeauthoryear{Muschaweck}{Muschaweck}{2022}]{Muschaweck2022}
Muschaweck, J.A. 2022, 9.
\newblock Tailored freeform surfaces for illumination with extended sources.
\newblock pp.\ ~9. SPIE-Intl Soc Optical Eng.
\newblock Presented at SPIE Optical Engineering + Applications, San Diego,
  California, 3 October 2022.

\bibitem[\protect\citeauthoryear{Nimier-David, Vicini, Zeltner, and
  Jakob}{Nimier-David et~al.}{2019}]{Mitsuba2}
Nimier-David, M., D.~Vicini, T.~Zeltner, and W.~Jakob. 2019, 11.
\newblock Mitsuba 2: A retargetable forward and inverse renderer.
\newblock {\em ACM Transactions on Graphics\/}~38: 1--17.
\newblock \doi{10.1145/3355089.3356498} .

\bibitem[\protect\citeauthoryear{Oertmann}{Oertmann}{1989}]{oertmann_differential_1989}
Oertmann, F.W. 1989.
\newblock Differential ray tracing formulae; applications especially to
  aspheric optical systems.
\newblock In {\em Optical Design Methods, Applications and Large Optics},
  Volume 1013, Hamburg, Germany, pp.\  20--26. SPIE.
\newblock Presented at 1988 International Congress on Optical Science and
  Engineering, Hamburg, Germany, 13 April 1989.

\bibitem[\protect\citeauthoryear{Pharr, Jakob, and Humphreys}{Pharr
  et~al.}{2016}]{pharr2016physically}
Pharr, M., W.~Jakob, and G.~Humphreys. 2016.
\newblock {\em Physically based rendering: From theory to implementation}.
\newblock Burlington, Massachusetts: Morgan Kaufmann Publishers.

\bibitem[\protect\citeauthoryear{Piegl and Tiller}{Piegl and
  Tiller}{1996}]{Piegl1995}
Piegl, L. and W.~Tiller. 1996.
\newblock {\em The NURBS book}.
\newblock Springer Science \& Business Media.

\bibitem[\protect\citeauthoryear{Schlick}{Schlick}{1994}]{Schlick1994}
Schlick, C. 1994, 8.
\newblock An inexpensive brdf model for physically-based rendering.
\newblock {\em Computer Graphics Forum\/}~13: 233--246.
\newblock \doi{10.1111/1467-8659.1330233} .

\bibitem[\protect\citeauthoryear{Sorgato, Chaves, Thienpont, and Duerr}{Sorgato
  et~al.}{2019}]{sorgato_design_2019}
Sorgato, S., J.~Chaves, H.~Thienpont, and F.~Duerr. 2019.
\newblock Design of illumination optics with extended sources based on
  wavefront tailoring.
\newblock {\em Optica\/}~6: 966--971.
\newblock \doi{10.1364/OPTICA.6.000966} .

\bibitem[\protect\citeauthoryear{Stone and Forbes}{Stone and
  Forbes}{1997}]{stone_differential_1997}
Stone, B.D. and G.W. Forbes. 1997, October.
\newblock Differential ray tracing in inhomogeneous media.
\newblock {\em Journal of the Optical Society of America A\/}~{\em 14\/}(10):
  2824.
\newblock \doi{10.1364/JOSAA.14.002824} .

\bibitem[\protect\citeauthoryear{Sun, Wang, Fu, Dun, and Heidrich}{Sun
  et~al.}{2021}]{Sun2021}
Sun, Q., C.~Wang, Q.~Fu, X.~Dun, and W.~Heidrich. 2021, 7.
\newblock End-to-end complex lens design with differentiate ray tracing.
\newblock {\em ACM Transactions on Graphics\/}~40: 1--13.
\newblock \doi{10.1145/3450626.3459674} .

\bibitem[\protect\citeauthoryear{Synopsis}{Synopsis}{2023}]{LightTools}
Synopsis. 2023.
\newblock Lighttools.
\newblock https://www.synopsys.com/optical-solutions/lighttools.html.

\bibitem[\protect\citeauthoryear{Synopsys}{Synopsys}{2023}]{codev}
Synopsys. 2023.
\newblock Code v.
\newblock https://www.synopsys.com/optical-solutions/codev.html.

\bibitem[\protect\citeauthoryear{Tukker}{Tukker}{2007}]{tukker_efficient_2007}
Tukker, T.W. 2007.
\newblock Efficient collimator design for extended light sources with the flux
  tube method.
\newblock Presented at SPIE Optical Engineering + Applications, San Diego,
  California, 18 September 2007.

\bibitem[\protect\citeauthoryear{Volatier, Álvaro Menduiña-Fernández, and
  Erhard}{Volatier et~al.}{2017}]{Volatier2017}
Volatier, J.B., Álvaro Menduiña-Fernández, and M.~Erhard. 2017, 7.
\newblock Generalization of differential ray tracing by automatic
  differentiation of computational graphs.
\newblock {\em Journal of the Optical Society of America A\/}~34: 1146.
\newblock \doi{10.1364/josaa.34.001146} .

\bibitem[\protect\citeauthoryear{Wu, Feng, Zheng, Liang, Benítez, Miñano, and
  Duerr}{Wu et~al.}{2018}]{Wu2018}
Wu, R., Z.~Feng, Z.~Zheng, R.~Liang, P.~Benítez, J.C. Miñano, and F.~Duerr.
  2018, 7.
\newblock Design of freeform illumination optics.
\newblock {\em Laser and Photonics Reviews\/}~12: 1700310.
\newblock \doi{10.1002/lpor.201700310} .

\end{thebibliography}
% %% if required, the content of .bbl file can be included here once bbl is generated
%%\input sn-article.bbl

%% Default %%
% \input sn-sample-bib.tex%

\end{document}